\documentclass{article} % For LaTeX2e
\usepackage{iclr2027_conference,times}

\usepackage{amsmath,amsfonts,bm}

\def\eqref#1{equation~\ref{#1}}
\def\1{\bm{1}}

\DeclareMathAlphabet{\mathsfit}{\encodingdefault}{\sfdefault}{m}{sl}
\SetMathAlphabet{\mathsfit}{bold}{\encodingdefault}{\sfdefault}{bx}{n}

\usepackage{hyperref}
\hypersetup{hidelinks}

\usepackage{url}

\usepackage{enumitem}
\usepackage{graphicx}
\usepackage{amsmath}
\usepackage{booktabs}
\usepackage{multirow}

\usepackage{makecell}
\usepackage{algorithm}
\usepackage{algpseudocode}
\usepackage{stmaryrd}

\title{DegreeSpar: Structured Degree Sparsity for Efficient Secure Transformer Inference}

\author{
Yifei Cai$^{1}$ \quad
Zhuoran Li$^{2}$ \quad
Xiaozuo Shen$^{2}$ \quad
Hongyi Wu$^{2}$ \quad
Chunsheng Xin$^{1}$ \\
$^{1}$Iowa State University, Ames, IA, USA \\
$^{2}$University of Arizona, Tucson, AZ, USA
}

\iclrfinalcopy % Uncomment for camera-ready version, but NOT for submission.
\begin{document}

\maketitle
\fancyhead[L]{}

\begin{abstract}
Secure Transformer inference protects sensitive inputs but incurs substantial cryptographic overhead, with nonlinear operations such as Softmax and GeLU becoming major bottlenecks. Existing compression methods reduce nonlinear complexity, sequence-dependent computation, or model structure through separately defined compression variables. Under aggressive compression, however, these independently optimized perturbations can accumulate: at a matched compression level, stacking representative approximation, token-pruning, and model-pruning methods reduces ViT-S accuracy from $80.20\%$ to $76.41\%$.
We introduce \textbf{DegreeSpar}, which formulates secure Transformer compression as \emph{structured sparsification over nonlinear polynomial degrees}. Polynomial degree directly controls the cost of secure nonlinear evaluation, while computation-aligned zero-degree structures expose token-level and model-dimension computation as removable within the same optimization space. 
DegreeSpar further incorporates approximation-aware training for low-degree Softmax and GeLU, enabling aggressive degree reduction and creating the optimization headroom required for structured computation removal.
Across vision and language Transformers, DegreeSpar consistently improves the accuracy--latency trade-off across model scales, tasks, and sequence lengths, achieving $2.29\times$--$6.63\times$ speedups over the corresponding baselines. Under the same network setting, DegreeSpar achieves $92.68\%$ accuracy on BERT/SST-2 in $110.55$\,s, compared with $92.66\%$ in $167.26$\,s for CipherPrune, the closest prior hybrid secure-inference approach. These results establish structured polynomial degree as an effective shared optimization space for secure Transformer compression.
\end{abstract}

\section{Introduction}
\label{Introduction}

\begin{figure*}[t]
\centering
\includegraphics[trim={0cm 0cm 0cm 0cm}, clip, scale=0.57]{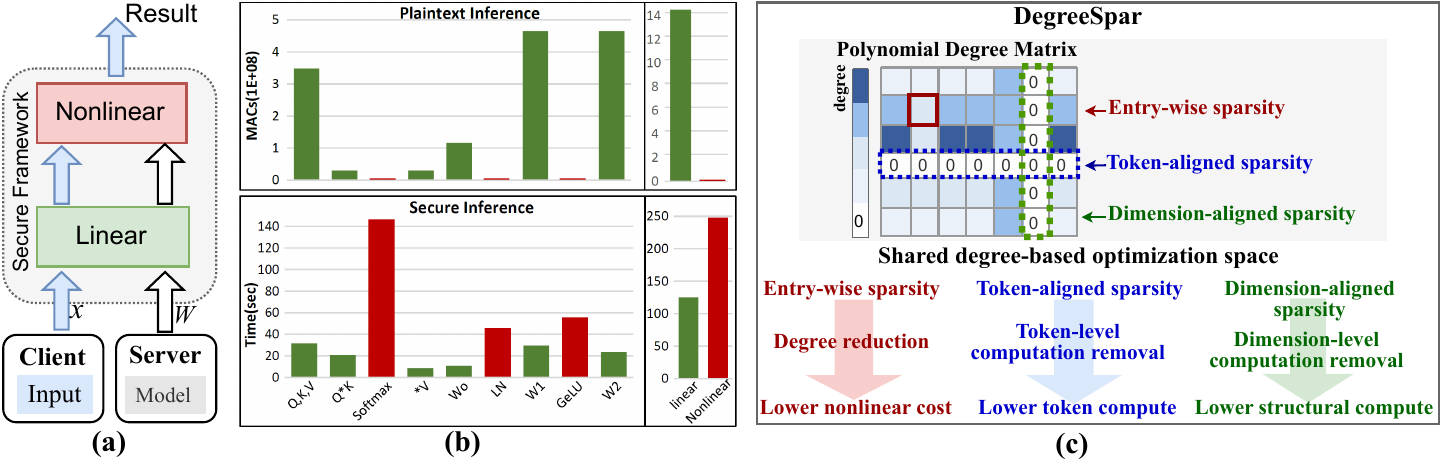}
\vspace*{-0.25in}
\caption{
(a) Secure Transformer inference.
(b) Secure execution shifts a substantial fraction of inference cost toward nonlinear operations under WAN1 (400 Mbps, 4 ms) network conditions.
(c) DegreeSpar organizes polynomial degrees into a shared structured optimization space. Entry-wise degree reduction lowers nonlinear cost, while token and dimension-aligned zero-degree sparsity enables the corresponding token-level and dimension-level computations to be removed.
}
\label{fig:f1}
\vspace*{-0.2in}
\end{figure*}

Transformers~\citep{vaswani2017attention} are increasingly deployed on third-party servers, making privacy-preserving inference important when user inputs are sensitive. Secure inference frameworks based on homomorphic encryption (HE) and secure multi-party computation (MPC), such as BOLT~\citep{pang2023bolt} and BumbleBee~\citep{lu2023bumblebee}, protect private inputs but remain orders of magnitude slower than plaintext execution, as illustrated in Figure~\ref{fig:f1}(a).
A major source of this overhead is nonlinear computation.
While linear operations such as matrix multiplications can be supported relatively efficiently through HE-domain arithmetic and parallel execution on scalable server-side hardware~\citep{moon2025thor}, nonlinear operations such as Softmax and GeLU require substantially more expensive secure evaluation, typically through look-up tables or polynomial approximation with communication-intensive interactive protocols~\citep{hao2022iron,zimerman2024converting}.
As shown in Figure~\ref{fig:f1}(b), under our WAN1 setting (400 Mbps, 4 ms), ViT-Base requires $373.11$\,s for secure inference, of which approximately $250$\,s are spent on nonlinear operations.
Efficient secure Transformer inference therefore requires compression strategies that explicitly target the dominant nonlinear computation while also reducing the substantial cost of linear operations.

Existing compression methods address different sources of secure-inference overhead through complementary mechanisms. Polynomial approximation reduces nonlinear evaluation complexity~\citep{kim2021bert,zimerman2024power}; token pruning reduces sequence-dependent computation~\citep{goyal2020power,liang2022not,rao2021dynamicvit}; and model pruning removes redundant structural computation~\citep{yu2022width}.
Despite their complementary benefits, these methods typically define compression in separate optimization spaces, even though their resulting computation and approximation errors interact within the same Transformer. This separation limits their ability to coordinate compression under aggressive cost reduction. For example, CipherPrune~\citep{zhang2025cipherprune}, the closest prior hybrid approach, couples token pruning with polynomial approximation through token-importance scores. However, low token importance does not necessarily imply that the associated nonlinear computation can tolerate a lower approximation degree.
The challenge becomes evident when independently combining multiple compression mechanisms. At a matched compression level, stacking representative approximation, token-pruning, and model-pruning methods reduces ViT-S accuracy from $80.20\%$ to $76.41\%$.
These observations raise a central question:
\textbf{Can nonlinear approximation, token removal, and model-dimension removal be coordinated through a shared optimization space that directly reflects secure-computation cost?}

Our key observation is that secure nonlinear computation already provides such a variable: the \textbf{polynomial degree}. Unlike conventional compression variables defined separately for nonlinear functions, tokens, and model dimensions, polynomial degree directly determines the cryptographic cost of nonlinear evaluation. Moreover, degree variables can be organized according to the computational structure of the Transformer, connecting nonlinear approximation with larger computational units.
Based on this observation, we introduce \textbf{DegreeSpar}, which formulates secure Transformer compression as \textbf{structured degree sparsification}. DegreeSpar organizes nonlinear evaluation degrees into a shared optimization space and learns sparsity at different computational granularities. Reducing individual degrees directly lowers the cost of nonlinear evaluation. By jointly optimizing computation-aligned degree groups and their associated structural parameters, DegreeSpar further exposes removable token- and dimension-level computation.
\textbf{The central idea is to optimize computation redundancy through a shared degree space, rather than independently designing compression mechanisms for different computational components.} Nonlinear approximation and structured computation removal consequently become coordinated outcomes of the same optimization process.

Realizing this formulation requires operating in an aggressively low-degree regime.
However, very-low-degree Softmax and GeLU approximations introduce substantial numerical error, preventing standard fine-tuning from fully recovering model accuracy and limiting how far degree variables can be sparsified.
DegreeSpar therefore incorporates \textbf{approximation-aware training} that explicitly adapts the model to the characteristic errors of low-degree nonlinear functions.
Adaptive noise injection exposes the model to perturbations concentrated in high-error regions of the low-degree approximation, while soft-boundary training smooths the abrupt transitions introduced by segmented approximations.
For example, on ViT-Base, directly using degree-2 approximations reduces accuracy from $81.40\%$ to $78.12\%$, whereas approximation-aware training recovers it to $81.28\%$ under the same approximation degree.
This robustness provides the optimization headroom required for more aggressive degree sparsification and subsequent structured computation removal.

DegreeSpar consequently achieves a stronger accuracy--latency trade-off than both stacked modular compression and existing secure-inference approaches.
Across vision and language Transformers with different model scales, tasks, and sequence lengths, DegreeSpar achieves $2.29\times$--$6.63\times$ end-to-end speedups over the corresponding uncompressed baselines.
At a matched compression level, DegreeSpar retains $80.12\%$ ViT-S accuracy, compared with $76.41\%$ for independently stacked approximation, token-pruning, and model-pruning methods.
We further compare against CipherPrune~\citep{zhang2025cipherprune}.
Under the same setting on BERT/SST-2, DegreeSpar achieves $92.68\%$ accuracy in $110.55$s, compared with $92.66\%$ in $167.26$s for CipherPrune.
These results show that coordinating compression through structured polynomial degrees provides substantial efficiency gains without the accuracy degradation caused by independently optimized compression decisions.

Our contributions are:
\vspace*{-0.10in}
\begin{itemize}[leftmargin=0in, itemindent=0.15in]
    \item We introduce \textbf{structured degree sparsity} for secure Transformer compression.
    By using nonlinear polynomial degree as a shared computation-aware optimization coordinate, DegreeSpar connects nonlinear cost reduction with token- and model-dimension computation removal through entry-wise, token-aligned, and dimension-aligned sparsity.

    \item We develop \textbf{DegreeSpar} to learn and instantiate this structure in practice.
    Approximation-aware training enables aggressively low polynomial degrees, while computation-aligned structured regularization induces removable zero-degree groups.
    Token-aligned sparsity is learned in a canonical rank space and securely instantiated under fixed per-layer retention budgets, whereas dimension-aligned sparsity is physically compacted into the deployed model offline.

    \item We demonstrate that degree-space optimization provides a stronger performance than existing compression approaches and 
    their independently stacked combinations.
    Across four Transformer configurations and five datasets, DegreeSpar achieves $2.29\times$--$6.63\times$ speedups.
    Under matched compression, it retains $80.12\%$ ViT-S accuracy compared with $76.41\%$ for modularly stacked approximation, token-pruning, and model-pruning methods; against CipherPrune, the closest hybrid approach, it reduces BERT/SST-2 latency from $167.26$\,s to $110.55$\,s at essentially matched accuracy.
    
\end{itemize}

\section{Background and Related Work}  % 缩短！
\label{sec:background}

\textbf{Secure Transformer Inference and Threat Model.}
Secure Transformer inference protects private inputs using cryptographic techniques such as homomorphic encryption (HE) and secure multi-party computation (MPC)~\citep{pang2023bolt,lu2023bumblebee}.
Following existing secure-inference systems~\citep{juvekar2018gazelle,279898,244032,rathee2020cryptflow2,pang2023bolt,lu2023bumblebee}, we consider the standard two-party semi-honest setting.
While linear operations can be supported by HE-based arithmetic, nonlinear operations such as Softmax and GeLU require substantially more expensive secure evaluation~\citep{hao2022iron} and are commonly implemented using polynomial representations~\citep{pang2023bolt,lu2023bumblebee}.
DegreeSpar builds on these primitives but treats polynomial degree as an optimization variable rather than a fixed approximation choice.
Detailed cryptographic definitions and security analysis are provided in Appendix~\ref{security_guarantee}.

\textbf{Transformer Compression.}
Transformer computation can be reduced through model pruning~\citep{yu2022width}, token pruning or merging~\citep{goyal2020power,liang2022not,rao2021dynamicvit,bolya2022token}, and nonlinear approximation~\citep{kim2021bert,zimerman2024power}.
CipherPrune~\citep{zhang2025cipherprune}, the closest prior hybrid approach, combines token pruning with polynomial approximation by using token-importance scores to determine pruning and indirectly guide lower-degree approximation.
However, these compression decisions remain separately parameterized or externally coupled, and their perturbations can accumulate under aggressive compression.
DegreeSpar instead derives degree reduction, token-aligned sparsity, and dimension-aligned sparsity from the same structured degree space.

\textbf{Secure Compression Execution.}
DegreeSpar determines model-dimension compression and per-layer token-retention budgets offline.
During inference, only the identities of retained tokens remain input dependent; they are selected through oblivious sorting and routing so that token scores, rankings, and pruning decisions remain hidden, while the observable tensor shapes are fixed by the public retention budgets.
Section~\ref{sec:structured_sparsification} describes how this fixed-policy execution is integrated with structured degree sparsity, and Appendices~\ref{Secure_token_drop} and~\ref{security_guarantee} provide the complete protocols and security analysis.

\clearpage

\section{DegreeSpar: Structured Degree Sparsification}
\label{sec:degreespar}

\begin{figure}[t]
\centering
\includegraphics[trim={0cm 0cm 0cm 0cm}, clip, scale=0.62]{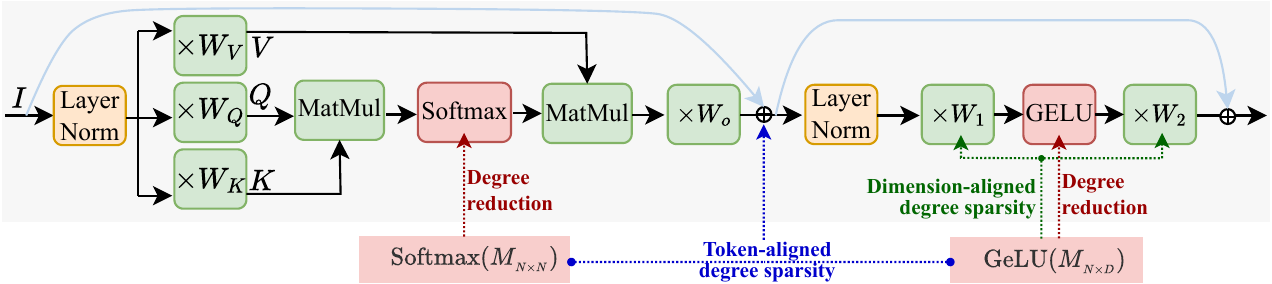}
\vspace*{-0.1in}
\caption{Overview of DegreeSpar within a Transformer layer.
Softmax and GeLU are represented by structured polynomial-degree matrices, $M_{N\times N}$ and $M_{N\times D}$.
Entry-wise degree reduction lowers nonlinear evaluation cost, while token-aligned and dimension-aligned degree sparsity expose removable token-level and dimension-level computation, respectively.}
\label{fig:outline}
\vspace*{-0.12in}
\end{figure}

\subsection{Structured Degree Space for Joint Compression}
\label{sec:degree_space}

DegreeSpar organizes secure Transformer compression in a shared structured optimization space over nonlinear secure-evaluation complexity. Rather than separately parameterizing nonlinear approximation, token pruning, and model pruning, DegreeSpar represents nonlinear computation through structured degree matrices and exploits different sparsity patterns over the same representation.
Illustrated in Figure~\ref{fig:outline}, we represent the nonlinear evaluations of Softmax and GeLU using degree matrices
$M^{S}\in\mathbb{R}^{N\times N}$ and
$M^{G}\in\mathbb{R}^{N\times D}$,
where $N$ is the sequence length and $D$ is the intermediate model dimension.

We use the term \emph{secure evaluation degree} as a unified abstraction of nonlinear secure-evaluation complexity.
For GeLU, it corresponds to the polynomial order; for Softmax, it corresponds to the repeated-squaring depth $n$ of the polynomial exponential approximation.
Reducing either eliminates expensive secure multiplication or square operations, making the evaluation degree a direct cost-aware optimization variable.
For brevity, we refer to these evaluation degrees simply as \emph{degrees} throughout the paper.

The structured degree representation connects computation reduction at two granularities. At the entry level, reducing a nonzero degree directly lowers the cost of the corresponding nonlinear evaluation. For Softmax, a lower repeated-squaring depth reduces secure square operations; for GeLU, eliminating higher-order polynomial terms reduces secure multiplication and square operations.
At the structural level, DegreeSpar organizes degree variables and their associated structural parameters into computation-aligned groups. When the learned sparsity eliminates the contribution of a token or an intermediate model dimension, the corresponding nonlinear and linear computation can be removed together.
This relationship allows DegreeSpar to extend fine-grained nonlinear cost reduction to larger-granularity computation removal within the same optimization space. The exact polynomial parameterization and its mapping to secure evaluation cost are detailed in Appendix~\ref{apdx:degree_cost}.

\subsection{Making Aggressive Degree Reduction Trainable}
\label{sec:approx_aware}

The effectiveness of structured degree sparsification depends on how far nonlinear evaluation degrees can be reduced while preserving model accuracy. Although lowering these degrees directly reduces secure-computation cost, aggressive approximation can introduce substantial numerical errors that limit further compression.

DegreeSpar addresses this challenge through \textbf{approximation-aware training}, which improves the model's tolerance to the characteristic errors of low-degree Softmax and GeLU approximations. This enables the model to operate in a more aggressively compressed degree regime, providing the foundation for subsequent structured sparsification.

For Softmax, reducing the repeated-squaring depth produces strongly non-uniform approximation error across the shifted-logit range.
We therefore use \textbf{adaptive noise injection} during offline training, concentrating perturbations in regions where the low-degree approximation exhibits large error.
The injected noise is not intended to reproduce the exact approximation error sample by sample; instead, it exposes the model to perturbations concentrated in the same high-error regions and improves its tolerance to aggressive Softmax degree reduction.
Noise injection is disabled during inference and introduces no additional online cost.

For GeLU, aggressively reducing the polynomial order introduces another optimization challenge.
The segmented approximation contains hard boundaries between polynomial regions, where low-order approximations can create abrupt changes in approximation error and gradients.
We therefore use \textbf{soft-boundary training}, replacing the hard transitions with smooth blending during offline optimization.
This relaxation stabilizes training near the segment boundaries and improves the model's tolerance to low-order GeLU approximation.
The original hard segmented function is restored for secure inference, so soft-boundary training introduces no additional online cost.

Together, these training mechanisms make aggressive degree reduction compatible with high model accuracy. By extending the range of usable low-degree configurations, approximation-aware training provides the optimization headroom needed to induce further computation-aligned sparsity.
Detailed algorithms and approximation-error analyses are provided in Appendix~\ref{apdx_Robust_Approx}.

\subsection{Computation-Aligned Structured Degree Sparsification}
\label{sec:structured_sparsification}

Entry-wise degree reduction lowers the cost of individual nonlinear evaluations, but reducing isolated operations does not necessarily eliminate the larger computational structures in which they participate.
DegreeSpar extends degree reduction to structured computation removal by aligning degree variables with the token- and dimension-level computation they collectively support. Through joint optimization of these computation-aligned groups and their associated structural parameters, DegreeSpar learns sparsity patterns that enable entire computational units to be removed.
We use three complementary regularization mechanisms.
\textbf{Staged $\ell_1/\ell_2$ regularization} progressively lowers individual secure evaluation degrees, which may remain at lower nonzero levels or reach zero.
\textbf{Mixed $\ell_1/\ell_2$ group regularization} further encourages variables aligned with the same token or model dimension to approach zero jointly.
For token-aligned sparsification, an associated structural coefficient $m_i\in[0,1]$ is optimized alongside the degree
variables, and a \textbf{binary surrogate} drives it toward an executable retain/drop state:
\begin{equation}
r_{\mathrm{grp}}
=
\sum_{g\in\mathcal{G}}\|M_g\|_2,
\qquad
r_{0\text{--}1}(m_i)
=
\big(m_i(1-m_i)\big)^2 .
\label{eq:structured_regularizers}
\vspace*{-0.1in}
\end{equation}

DegreeSpar progressively sparsifies the same degree space at
different computational granularities.
Entry-wise regularization first lowers individual evaluation
degrees to reduce nonlinear cost.
As the degrees approach low-order states, computation-aligned
group regularization is introduced to encourage removable
zero structures at the token and model-dimension levels.
The binary surrogate further encourages discrete
token-retention decisions.
Detailed schedules and coefficients are provided in
Appendix~\ref{Regularization_Setting}.
The learned sparsity patterns are converted into removable computation structures during offline compression, and the resulting compressed model is validated before deployment.

\begin{figure}[h]
\centering
\includegraphics[trim={0cm 0cm 0cm 0cm}, clip, scale=0.77]{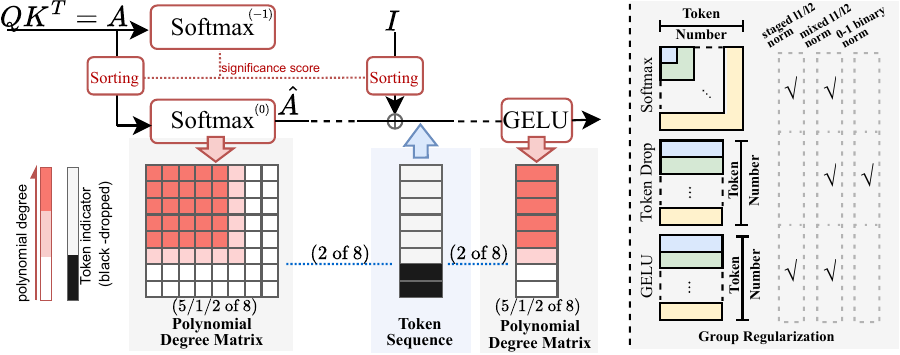}
\vspace*{-0.1in}
\caption{
Token-aligned joint structured degree optimization.
Tokens are aligned with their corresponding Softmax and GeLU degree variables according to significance scores. Computation-aligned group regularization jointly reduces these degree variables and induces removable token-aligned sparsity.
}
\label{fig:token_joint}
\vspace*{-0.10in}
\end{figure}

\paragraph{Token-aligned degree groups.}
Token-level redundancy is inherently input dependent: a token that is important for one input may be redundant for another.
Directly optimizing sparsity over absolute token identities would therefore produce structures that do not transfer reliably across inputs.

DegreeSpar instead maps input-dependent token identities into a common \emph{rank space}.
For each input, tokens are sorted according to their significance scores, and the same permutation is applied to the associated Softmax and GeLU degree variables, as illustrated in Figure~\ref{fig:token_joint}.
Structured regularization is then applied to these rank-aligned groups.
The degree variables are optimized together with the associated token-retention coefficient, encouraging each rank-aligned
computation group toward a removable state.
A token can be removed once the learned sparsity eliminates its contribution, including when the effective GeLU coefficient
or the token-retention coefficient reaches zero.
The associated Softmax computation can then be skipped withoutrequiring its degree variables to reach zero independently.

Importantly, the significance score determines which token occupies each canonical rank position, but does not directly prescribe the secure evaluation degree of that position or whether it is ultimately removed.
These compression decisions are learned through structured degree sparsification.
Thus, token importance provides a transferable coordinate system for optimization rather than serving as a direct proxy for nonlinear approximation tolerance.

We use \textbf{Variance-Guided CLS Attention (VCA)} as the significance estimator.
VCA emphasizes token-wise feature variance in shallow layers and progressively shifts toward \texttt{[CLS]} attention in deeper layers.
Both signals are derived from statistics already available within the Transformer, avoiding an auxiliary prediction network in the encrypted domain.
Its formulation and comparison with attention-based scoring are provided in Appendix~\ref{VCA}.

After offline optimization, DegreeSpar fixes the per-layer retention budget $K_l$ and the learned degree configurations
for different token-importance ranges.
For each private input, secure VCA-based ranking maps tokens to these predefined ranges, where they are evaluated using
the corresponding fixed degree configurations.

Thus, DegreeSpar learns a fixed compression policy in rank space, while allowing the identities of retained tokens
to vary across inputs.
The secure realization of this policy is described below.

\begin{figure}[h]
\centering
\includegraphics[trim={0cm 0cm 0cm 0cm}, clip, scale=0.79]{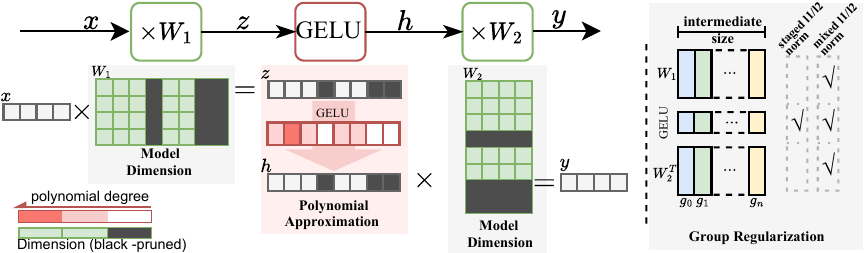}
\vspace*{-0.1in}
\caption{
Dimension-aligned joint structured degree optimization.
GeLU degree variables are aligned with their associated model dimensions and surrounding linear computation. The resulting degree-aligned sparsity enables the associated model weights and intermediate-dimension computation to be removed together.
}
\label{fig:dimension_joint}
\vspace*{-0.10in}
\end{figure}

\paragraph{Dimension-aligned degree groups.}
The same structured degree mechanism applies along the model-dimension axis.
Consider an FFN block,
$y=\mathrm{GeLU}(xW_1)W_2$.
Each intermediate dimension jointly corresponds to one output dimension of $W_1$, the associated GeLU computation, and the matching input dimension of $W_2$.

As illustrated in Figure~\ref{fig:dimension_joint}, DegreeSpar groups the GeLU degree variables associated with the same intermediate dimension.
Equivalently, in the GeLU degree matrix
$M^{G}\in\mathbb{R}^{N\times D}$,
each dimension-aligned group corresponds to one column.
Structured regularization encourages the degrees within each group to decrease jointly, while the associated dimensions of $W_1$ and $W_2$ are optimized consistently with the learned degree structure.
Once the resulting sparsity eliminates an intermediate dimension's contribution, its associated GeLU and linear computation can be removed together.

Unlike token identities, model dimensions are input independent.
The resulting sparse dimensions are therefore physically removed \emph{offline} before secure deployment: the corresponding output dimension of $W_1$, GeLU coordinate, and input dimension of $W_2$ are jointly removed, and the remaining weights are compacted into smaller dense matrices.
Online inference directly evaluates this fixed compacted model without any input-dependent dimension-pruning decision.
The compaction procedure is detailed in Appendix~\ref{apdx:dimension_compaction}.

\paragraph{Secure fixed-policy execution.}
The learned token- and dimension-level sparsity structures are instantiated differently at deployment.
Dimension-aligned sparsity is compiled into the model offline: sparse dimensions are physically removed and secure inference directly evaluates the resulting compacted model.
For token-aligned sparsity, the per-layer retention budget $K_l$
and the degree configurations for different rank ranges are
determined offline.
During secure inference, private token rankings determine which
tokens are retained and which predefined degree configuration
is applied to each token.

During secure inference, token significance scores and rankings remain protected within the secure-computation domain.
Oblivious sorting maps private token identities into the learned rank space, after which the fixed top-$K_l$ budget is applied without revealing which tokens are retained or removed.
Thus, the observable sequence length is fixed across inputs, while input-dependent token selection remains private.
The complete oblivious ranking, selection, and routing protocols are provided in Appendix~\ref{Secure_token_drop}, and the security analysis is given in Appendix~\ref{security_guarantee}.

% \clearpage
\section{Evaluation}
\label{Evaluation}

\subsection{Experimental Setup}
\label{Experimental_Setup}

\textbf{Models and datasets.}
We evaluate DegreeSpar on ViT-Small, ViT-Base, and ViT-Large~\citep{dosovitskiy2020image} using ImageNet-1K~\citep{krizhevsky2012imagenet}, and BERT-Base~\citep{devlin2019bert} using IMDB~\citep{IMDB} and three GLUE tasks~\citep{GLUE}: CoLA, SST-2, and MNLI. These experiments cover different model scales, tasks, and sequence lengths ranging from 64 to 512 tokens. Detailed training configurations are provided in Appendix~\ref{Training_Setting}.

\textbf{Secure inference.}
We implement DegreeSpar on secure framework BumbleBee~\citep{lu2023bumblebee}, built on the SPU library~\citep{spu}, and measure end-to-end secure inference latency on an AMD 3995 64-core CPU testbed. We evaluate three network conditions: \textbf{LAN} (3~Gbps, 0.8~ms), \textbf{WAN1} (400~Mbps, 4~ms), and \textbf{WAN2} (100~Mbps, 10~ms), following prior secure-inference evaluations~\citep{lu2023bumblebee,pang2023bolt,zhang2024secure}. Unless otherwise specified, we use the framework's default security and execution settings.

\subsection{Validating the Core Design of DegreeSpar}
\label{sec:core_validation}

We first validate the two principles underlying DegreeSpar: aggressive degree reduction can be made compatible with high model accuracy, and coordinating compression within a shared degree space preserves substantially more accuracy than independently combining compression methods under matched budgets.

\vspace*{-0.15in}
\begin{table}[h]
\centering
\caption{Validation of two key properties of DegreeSpar.
(a) Accuracy under the same degree-2 approximation with and without approximation-aware training.
(b) DegreeSpar versus independently stacked compression methods under a matched compression budget.}
\label{tab:core_validation}
\vspace{0.03in}
\small
\setlength{\tabcolsep}{5pt}

\begin{minipage}[t]{0.53\linewidth}
\centering
\textbf{(a) Aggressive degree reduction}
\vspace{0.04in}

\begin{tabular}{lcc}
\toprule
\textbf{Training} &
\textbf{ViT-S} &
\textbf{ViT-B} \\
\midrule
Naive degree-2
& 79.38\% & 78.12\% \\
Approx-aware
& \textbf{80.13}\% & \textbf{81.28}\% \\
\bottomrule
\end{tabular}
\end{minipage}
\hfill
\begin{minipage}[t]{0.44\linewidth}
\centering
\textbf{(b) Modular vs.\ degree-space compression}
\vspace{0.04in}

\begin{tabular}{lc}
\toprule
\textbf{Method} & \textbf{Acc.} \\
\midrule
Naive stacked & 76.41\% \\
DegreeSpar     & \textbf{80.12}\% \\
\bottomrule
\end{tabular}
\end{minipage}

\vspace{-0.08in}
\end{table}

\textbf{Enabling aggressive degree reduction.}
As shown in Table~\ref{tab:core_validation}(a), proposed approximation-aware training improves ViT-S accuracy from 79.38\% to 80.13\% and ViT-B accuracy from 78.12\% to 81.28\% under the same degree-2 approximation.
These results demonstrate that approximation-aware training substantially improves the model's tolerance to low-degree nonlinear evaluation, enabling more aggressive degree reduction without the severe accuracy degradation observed with naive approximation.

\textbf{Shared degree-space optimization versus modular compression.}
We compare DegreeSpar with a modular combination of existing nonlinear approximation~\citep{zimerman2024power}, token pruning~\citep{liang2022not}, and model pruning~\citep{yu2022width}. We retain the original hyperparameter settings of these methods while matching their respective approximation degrees, token-retention budgets, and model-dimension pruning ratios to those of DegreeSpar.
As shown in Table~\ref{tab:core_validation}(b), the modular combination reduces ViT-S accuracy from 80.20\% to 76.41\%, whereas DegreeSpar retains 80.12\%.
This comparison demonstrates that the benefit of DegreeSpar extends beyond applying multiple compression mechanisms: coordinating their compression effects within a shared degree space enables substantially better accuracy preservation under matched compression levels.

\subsection{End-to-End Secure Inference Performance}
\label{sec:end_to_end}

\vspace{-0.1in}

% ============================================================
% Table 1 + Table 2: side-by-side direct comparisons
% ============================================================
\begin{table}[h]
\centering
\caption{End-to-end comparison with existing secure Transformer compression approaches.
(a) ViT-B/ImageNet-1K under different network conditions.
(b) BERT/SST-2 comparison with CipherPrune; CipherPrune* is configured to match DegreeSpar's latency.
Latency is reported in seconds.}
\label{tab:e2e_comparison}
\vspace{0.03in}
\small

\begin{minipage}[t]{0.58\linewidth}
\centering
\textbf{(a) ViT-Base / ImageNet-1K}
\vspace{0.04in}

\setlength{\tabcolsep}{3.8pt}
\begin{tabular}{lrrrr}
\toprule
\textbf{Method} &
\textbf{LAN} &
\textbf{WAN1} &
\textbf{WAN2} &
\textbf{Acc.} \\
\midrule
Baseline
    & 152.26 & 373.11 & 1144.43 & 81.40\% \\
WDPruning
    & 120.54 & 221.37 & 911.65 & 80.76\% \\
PowerSoftmax
    & 150.49 & 323.64 & 940.15 & 81.35\% \\
EViT
    & 93.68  & 297.39 & 665.59 & 81.13\% \\
\textbf{DegreeSpar}
    & \textbf{56.85} & \textbf{95.57}
    & \textbf{366.17} & \textbf{81.24}\% \\
\bottomrule
\end{tabular}
\end{minipage}
\hfill
\begin{minipage}[t]{0.39\linewidth}
\centering
\textbf{(b) BERT-Base / SST-2 (WAN2)}
\vspace{0.04in}

\setlength{\tabcolsep}{4pt}
\begin{tabular}{lrr}
\toprule
\textbf{Method} &
\textbf{Latency} &
\textbf{Acc.} \\
\midrule
CipherPrune
    & 167.26 & 92.66\% \\
CipherPrune*
    & 110.57 & 88.00\% \\
\textbf{DegreeSpar}
    & \textbf{110.55} & \textbf{92.68}\% \\
\bottomrule
\end{tabular}
\end{minipage}

\vspace{-0.08in}
\end{table}

\textbf{Comparison with existing approaches.}
We compare DegreeSpar with representative compression approaches, including WDPruning~\citep{yu2022width}, PowerSoftmax~\citep{zimerman2024power}, and EViT~\citep{liang2022not}.
We reproduce their published hyperparameters and compression configurations, and measure end-to-end latency on the same secure-inference platform under identical hardware and network conditions.
Detailed configurations are provided in Appendix~\ref{apdx:baseline_comparison}.

As shown in Table~\ref{tab:e2e_comparison}(a), DegreeSpar achieves lower end-to-end latency across all three network conditions while maintaining comparable accuracy.
On ViT-B under WAN2, DegreeSpar reduces inference latency from 1144.43\,s to 366.17\,s while retaining 81.24\% accuracy, compared with the baseline accuracy of 81.40\%.
We further compare DegreeSpar with CipherPrune~\citep{zhang2025cipherprune}, the closest prior hybrid secure-inference approach.
As shown in Table~\ref{tab:e2e_comparison}(b), DegreeSpar achieves 92.68\% accuracy on BERT/SST-2 in 110.55\,s, compared with 92.66\% accuracy in 167.26\,s for CipherPrune.
When CipherPrune is configured to match DegreeSpar's latency (denoted as CipherPrune*), its accuracy decreases to 88.00\%. These complementary comparisons demonstrate that DegreeSpar achieves a better accuracy--latency trade-off than CipherPrune in the evaluated setting.

% ============================================================
% Table 3: broad end-to-end evaluation
% ============================================================
\begin{table}[h]
\centering
\caption{End-to-end performance across model scales, tasks, and sequence lengths under WAN2. Performance is reported as MCC (\%) for CoLA and Accuracy (\%) for all other tasks. Latency is reported in seconds.}
\label{tab:broad_e2e}
\resizebox{134mm}{!}{
\begin{tabular}{ccc|cc|cc|c}
\Xhline{1.5pt}

\multirow{2}{*}{\textbf{Model}} &
\multirow{2}{*}{\textbf{Dataset}} &
\multirow{2}{*}{\textbf{Seq. Len.}} &
\multicolumn{2}{c|}{\textbf{Baseline}} &
\multicolumn{2}{c|}{\textbf{DegreeSpar}} &
\multirow{2}{*}{\textbf{Speedup}} \\

& & &
\textbf{Acc.} &
\textbf{Latency} &
\textbf{Acc.} &
\textbf{Latency} &
\\ \hline

ViT-Small &
ImageNet-1K &
197 &
80.20\% &
538.52 &
80.12\% &
132.02 &
$4.08\times$ \\

ViT-Base &
ImageNet-1K &
197 &
81.40\% &
1144.43 &
81.24\% &
366.17 &
$3.13\times$ \\

ViT-Large &
ImageNet-1K &
197 &
81.89\% &
3171.59 &
80.95\% &
1384.31 &
$2.29\times$ \\ \hline

BERT-Base &
IMDB &
197 &
89.90\% &
1144.43 &
88.78\% &
423.67 &
$2.70\times$ \\

BERT-Base &
CoLA &
64 &
58.80\% &
241.12 &
58.06\% &
58.64 &
$4.11\times$ \\

BERT-Base &
SST-2 &
128 &
91.74\% &
604.03 &
92.68\% &
110.55 &
$5.46\times$ \\

BERT-Base &
MNLI &
256 &
84.69\% &
1729.24 &
83.82\% &
260.96 &
$6.63\times$ \\

BERT-Base &
IMDB-Long &
512 &
92.65\% &
5577.23 &
91.40\% &
1035.15 &
$5.39\times$ \\ \hline

\Xhline{1.5pt}
\end{tabular}}
\end{table}

\textbf{Performance across models, tasks, and sequence lengths.}
Table~\ref{tab:broad_e2e} reports DegreeSpar's performance across four Transformer configurations, five datasets, and sequence lengths ranging from 64 to 512 tokens.
Under WAN2, DegreeSpar achieves $2.29\times$--$6.63\times$ end-to-end speedups while largely preserving task performance across all evaluated settings.
These results demonstrate that the efficiency gains extend across different model scales, tasks, and input lengths.

\subsection{Efficiency Breakdown and Learned Sparsity}
\label{sec:efficiency_breakdown}

\begin{figure}[t]
\centering
\includegraphics[trim={0cm 0cm 0cm 0cm}, clip, scale=0.52]{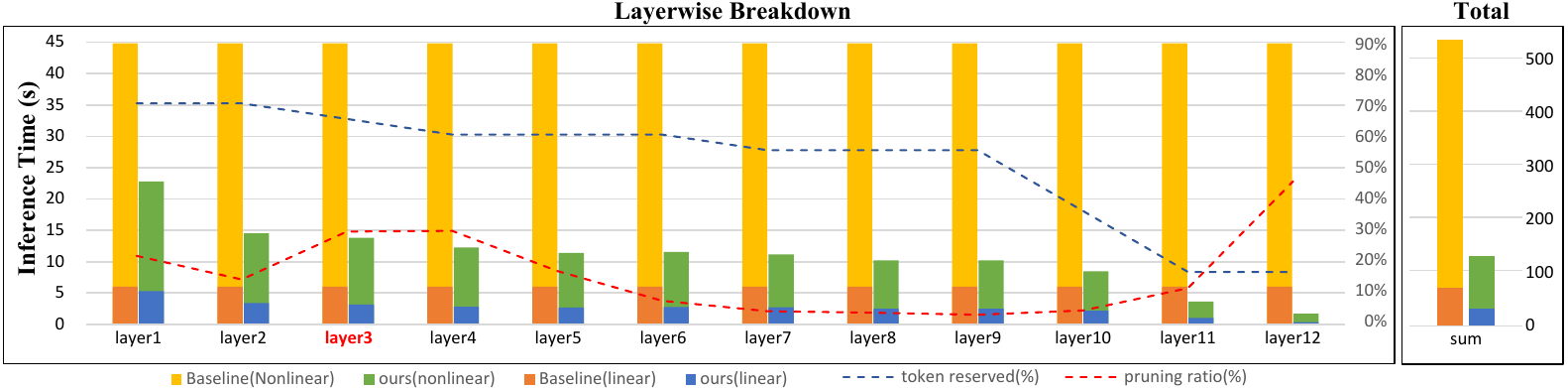}
\vspace{-0.2in}
\caption{Layer-wise inference-cost breakdown of DegreeSpar on ViT-S/ImageNet under WAN2.}
\label{fig:layerwise_breakdown}
\vspace{-0.10in}
\end{figure}

We further examine how learned degree sparsity translates into secure-inference cost reduction. Figure~\ref{fig:layerwise_breakdown} shows the layer-wise inference cost of ViT-S on ImageNet under WAN2. DegreeSpar reduces nonlinear evaluation costs through degree reduction, while learned token- and dimension-aligned sparsity further removes the redundancy associated with tokens and model dimensions.

\textbf{A closer look at Layer-3.}
Layer-3 illustrates how structured degree sparsification translates into computation reduction. DegreeSpar lowers the evaluation degrees of Softmax and GeLU, directly reducing their nonlinear evaluation costs. 
Moreover, the learned sparsity enables the removal of 69 out of 197 token-aligned groups and 436 out of 1536 dimension-aligned groups, corresponding to approximately 35\% token sparsity and 28\% model-dimension sparsity, respectively. Removing these computational structures reduces both linear and nonlinear costs.

\textbf{Sources of overall cost reduction.}
DegreeSpar jointly optimizes degree reduction and structured computation removal within a shared degree space, making their efficiency gains inherently coupled.
To characterize these gains, we sequentially account for the incremental cost reductions from entry-wise degree reduction, token-level removal, and dimension-level removal.
Under this evaluation order, they account for 42\%, 54\%, and 4\% of the total cost reduction on ViT-S/ImageNet-1K, respectively.
These order-dependent measurements reflect the additional savings at each stage rather than independent contributions; in particular, the 4\% attributed to dimension-level removal is measured after the preceding reductions have already eliminated substantial computation.

% \vspace{-0.1in}
\subsection{Robustness and practical overhead.}
% \vspace{-0.1in}

Across five independent training runs, DegreeSpar maintains stable accuracy, with standard deviations of 0.1--0.2 percentage points across the evaluated configurations.
Secure token selection introduces only 1.73\,s of overhead for a 197-token sequence under WAN2, accounting for 0.32\% of the uncompressed inference latency.
This overhead is included in all reported end-to-end results.
Additional analyses of training stability and hyperparameter sensitivity are provided in Appendix~\ref{apdx:robustness}.

\section{Conclusion}
\label{Conclusion}

We introduced \textbf{DegreeSpar}, which formulates secure Transformer compression as structured sparsification over a shared space of degree variables. By organizing these variables into computation-aligned structures, DegreeSpar connects fine-grained nonlinear cost reduction with token- and dimension-level computation removal within a unified optimization framework.
Across vision and language Transformers, DegreeSpar achieves $2.29\times$--$6.63\times$ end-to-end speedups while preserving task performance.
Under matched compression levels, it retains 80.12\% ViT-S accuracy compared with 76.41\% for independently stacked compression methods.
These results show that coordinating compression through a shared degree space can substantially reduce secure-inference cost while avoiding the accuracy degradation observed when separate compression mechanisms are independently combined.

\clearpage
% \subsection*{AI Use Statement}
% % \vspace{-0.1in}
% Generative AI tools were used to assist with literature search and summarization, manuscript organization and restructuring, drafting and revising portions of the manuscript, and language polishing. The core research ideas, methodological contributions, experimental designs, and scientific conclusions were developed and validated by the authors.

% \clearpage
\bibliography{iclr2027_conference}
\bibliographystyle{iclr2027_conference}

\appendix

% ============================================================
% Appendix: Technical Details of DegreeSpar
% ============================================================

\section{Secure Evaluation Degree Parameterization and Secure Cost}
\label{apdx:degree_cost}

DegreeSpar uses the evaluation degree of each secure nonlinear
evaluation as its basic compression coordinate.
This appendix gives the concrete polynomial parameterization and
explains how reducing a degree variable translates into lower
secure-inference cost.

\subsection{Polynomial Parameterization}

For Softmax, we follow existing secure Transformer inference
frameworks and approximate the exponential component as
\begin{align}
\mathrm{Softmax}(x)_{ij}
&=
\frac{e^{x_{ij}}}
{\sum_{j\in[m]} e^{x_{ij}}},
\\
e^{\tilde{x}}
&\approx
\left(
1+\frac{\tilde{x}}{2^n}
\right)^{2^n},
\qquad
\tilde{x}=x-x_{\max}.
\label{eq:apdx_softmax}
\end{align}
For numerical stability, values below a truncation threshold
$T_{\exp}$ are treated as negligible.
The parameter $n$ controls the repeated-squaring depth and therefore
directly controls the secure evaluation cost
~\citep{pang2023bolt,lu2023bumblebee}.

For GeLU, secure inference uses a segmented polynomial
approximation~\citep{hao2022iron,pang2023bolt,lu2023bumblebee}.
A representative fourth-order form is
\begin{align}
\mathrm{ApproxGeLU}(x)=
\begin{cases}
x, & x>B, \\[2pt]
a|x|^4+b|x|^3+c|x|^2+d|x|+e+0.5x,
& |x|\le B, \\[2pt]
0, & x<-B,
\end{cases}
\label{eq:apdx_gelu}
\end{align}
where $B$ specifies the segment boundary. 
BumbleBee~\citep{lu2023bumblebee} employs a four-segment polynomial approximation of GeLU with a maximum polynomial order of six. In contrast, we adopt this three-segment fourth-order approximation~\citep{pang2023bolt} as our baseline and further optimize its polynomial degrees.

DegreeSpar organizes these nonlinear degrees into the structured
degree matrices described in Section~\ref{sec:degree_space}.
During offline optimization, degree regularization progressively
pushes individual nonlinear evaluations toward lower-degree
states. Importantly, these states are not predefined
token categories. The learned degree matrix can contain different
degree levels across nonlinear evaluations; a state such as
$n=2$ or $n=0$ observed in a particular layer is an execution
snapshot of this optimization rather than a predefined grouping.

Reducing nonzero degrees lowers the cost of individual nonlinear evaluations, whereas computation-aligned sparsity enables larger computational units to be removed. The associated structural parameters are optimized consistently with the learned degree structure, and structural removal does not require every degree variable in a group to reach zero.

\subsection{Why Degree Reduction Lowers Secure Cost}

The exponential approximation in Softmax is dominated by repeated
secure squaring. Reducing $n$ decreases the number and depth of
these secure square operations.
For GeLU, the polynomial order determines the required secure
multiplications and squares. Lower-order polynomials therefore
reduce both multiplicative depth and communication-intensive
nonlinear operations.

Table~\ref{tab:apdx_approx_cost} reports the secure computational
cost of different approximation degrees under our secure-inference
configuration.
For Softmax, reducing the square depth from $n=6$ to $n=2$
reduces the measured cost from $34.79$s to $13.32$s.
For GeLU, reducing a sixth-order polynomial to a second-order
polynomial reduces the corresponding cost from $31.01$s to $7.60$s.
Because secure multiplication is more expensive than secure
squaring in our backend, eliminating multiplication terms can
produce particularly large reductions.

\begin{table}[h]
\centering
\caption{
Secure computational cost of Softmax exponential and GeLU
approximations at different degree levels under our
secure-inference configuration.
}
\vspace*{0.05in}
\label{tab:apdx_approx_cost}
\small
\resizebox{\textwidth}{!}{
\begin{tabular}{c|ccc|c l c|ccccc|c}
\cline{1-5} \cline{7-13}

\textbf{Approx-EXP}
& \#Seg.
& \texttt{f\_less}
& \texttt{f\_square}
& \textbf{Total}
&
&
\textbf{Approx-GeLU}
& \#Seg.
& \#Degree
& \texttt{f\_less}
& \texttt{f\_square}
& \texttt{f\_mul}
& \textbf{Total}
\\

\cline{1-5} \cline{7-13}

$n=6$
& 2 & 2.58 & 32.21 & 34.79
&
&
poly-6
& 4 & 6 & 6.31 & 6.91 & 17.79 & 31.01
\\

$n=5$
& 2 & 2.58 & 29.70 & 32.28
&
&
poly-5
& 3 & 5 & 4.14 & 6.91 & 11.86 & 24.61
\\

$n=4$
& 2 & 2.58 & 23.19 & 25.77
&
&
poly-4
& 3 & 4 & 4.14 & 6.91 & 5.93 & 16.98
\\

$n=3$
& 2 & 2.58 & 16.10 & 18.68
&
&
poly-3
& 3 & 3 & 4.14 & 3.46 & 5.93 & 15.15
\\

$n=2$
& 2 & 2.58 & 10.74 & 13.32
&
&
poly-2
& 3 & 2 & 4.14 & 3.46 & 0.00 & 7.60
\\

$n=1$
& 2 & 2.58 & 5.94 & 8.52
&
&
poly-1
& 2 & 1 & 2.05 & 0.00 & 0.00 & 2.05
\\

\cline{1-5} \cline{7-13}
\end{tabular}}
\end{table}

This cost mapping is the reason evaluation degree is particularly
useful as the optimization variable for secure inference:
unlike an abstract pruning score, changing the degree directly
changes the cost of the underlying cryptographic computation.

% ============================================================
\section{Approximation-Aware Training}
\label{apdx_Robust_Approx}

Very-low-degree approximation substantially increases numerical
error. Approximation-aware training is therefore used during
offline optimization to increase the model's tolerance to the
specific error distribution introduced by aggressive degree
reduction. The auxiliary mechanisms described below are used
only during training; deployment uses the learned polynomial
approximations directly.

\subsection{Approximation Error under Aggressive Degree Reduction}

Reducing Softmax to a low square depth and GeLU to a low polynomial
order produces substantially larger approximation errors than
the higher-degree secure approximations normally used in
Transformer inference.
The error is also nonuniform: some input regions exhibit much
larger error than others, and segmented GeLU approximation
introduces abrupt changes around its approximation boundaries.

\begin{figure}[h]
\centering
\includegraphics[width=0.99\linewidth]{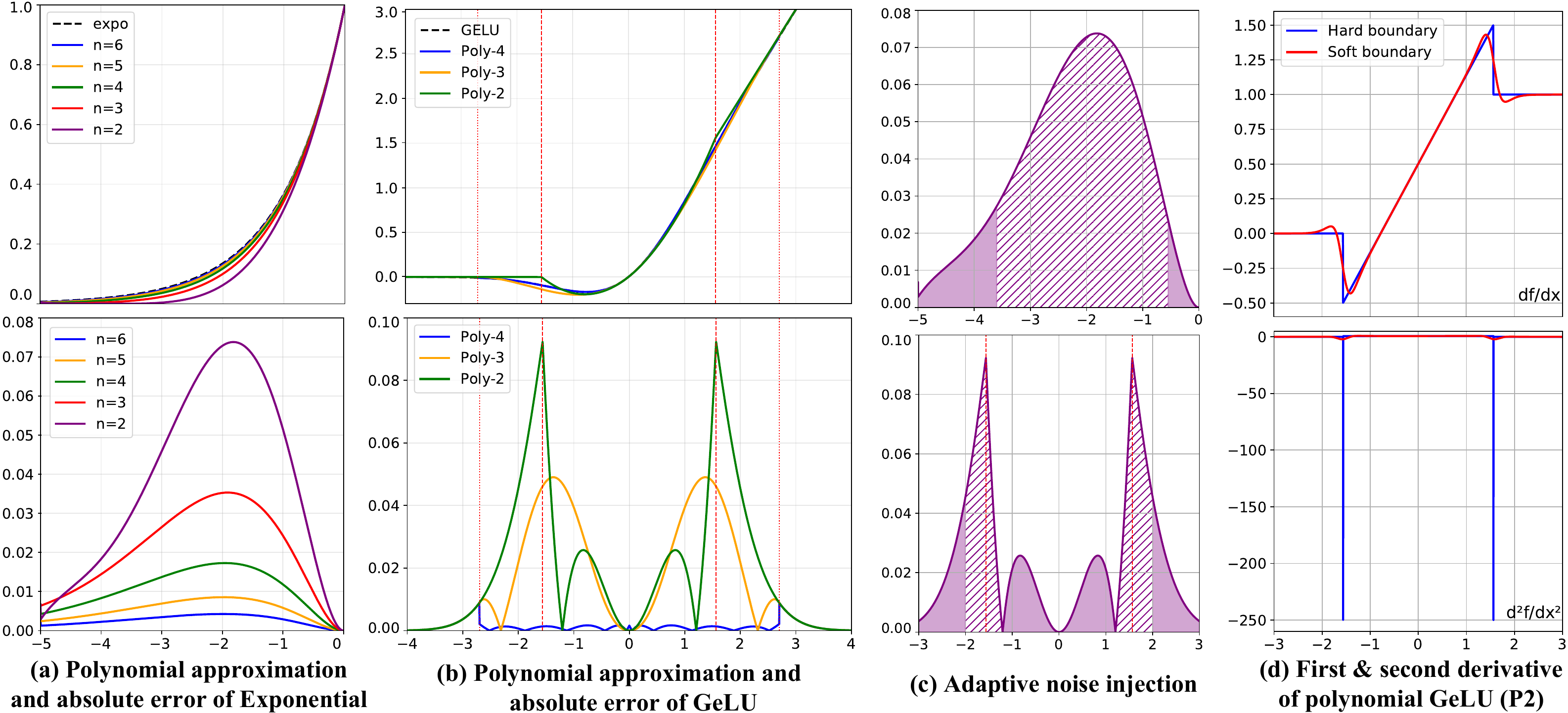}
\vspace{-0.1in}
\caption{
Approximation behavior of low-degree Softmax and GeLU and the
training-time mechanisms used by approximation-aware training.
}
\label{fig:apdx_approx_cmp}
\end{figure}

\subsection{Adaptive Noise Injection for Softmax}

For Softmax, we inject perturbations specifically into the regions
where the low-degree approximation exhibits large error.
The purpose is not to reproduce the exact approximation error
sample by sample, but to expose the model to perturbations with
the same characteristic high-error regions during offline
fine-tuning.

Given the shifted logits
\begin{equation}
    \mathbf z = \mathbf x-x_{\max},
\end{equation}
we sample
\begin{equation}
    \epsilon \sim \mathcal U(-\eta,\eta)
\end{equation}
for entries satisfying
$z_i\in\mathcal R_{\mathrm{noise}}$.
Noise injection is disabled during inference.

\begin{algorithm}[h]
\caption{Approximation-Aware Softmax Training}
\label{alg:robust_softmax}
\begin{algorithmic}[1]
\Require input $\mathbf{x}$, square depth $n$,
threshold $T_{\exp}$,
noise region $\mathcal R_{\mathrm{noise}}$,
noise magnitude $\eta$
\State $\mathbf z \gets \mathbf x-\max(\mathbf x)$
\If{training}
    \For{each $z_i \in \mathcal R_{\mathrm{noise}}$}
        \State $z_i \gets z_i+\epsilon$,
        $\epsilon\sim\mathcal U(-\eta,\eta)$
    \EndFor
\EndIf
\State $\mathbf y\gets
(1+\mathbf z/2^n)^{2^n}$
\State $\widetilde e_i\gets
    y_i$ if $z_i\ge T_{\exp}$,
    and $0$ otherwise
\State \Return
$\widetilde{\mathbf e}/
\sum_i\widetilde e_i$
\end{algorithmic}
\end{algorithm}

For the representative degree-2 Softmax configuration,
we use
\begin{equation}
\mathcal R_{\mathrm{noise}}
=
[-3.6,-0.55],
\qquad
T_{\exp}=-5.0,
\qquad
\eta=0.05.
\end{equation}
Adaptive noise injection forces the model to encounter
approximation-like perturbations during training and reduces its
reliance on fragile logit differences in high-error regions.

\subsection{Soft-Boundary Training for GeLU}

Besides noise injection, low-degree segmented GeLU introduces another optimization
problem. In addition to approximation error, hard transitions
between polynomial segments create abrupt variation around the
segment boundaries.
We therefore replace the hard boundary with a smooth transition
during training.

For the second-order central approximation,
\begin{equation}
m(x)=a|x|^2+b|x|+c+0.5x,
\end{equation}
we define
\begin{align}
w_{\mathrm{pos}}
&=
\sigma(k(x-B)),
\\
w_{\mathrm{neg}}
&=
\sigma(k(-x-B)),
\end{align}
and use the training-time softened function
\begin{equation}
y =
w_{\mathrm{pos}}x
+
\max
\left(
0,
1-w_{\mathrm{pos}}-w_{\mathrm{neg}}
\right)m(x).
\end{equation}
This smoothing is used only for optimization.
At deployment, DegreeSpar returns to the hard segmented
low-degree polynomial required by secure inference.

\begin{algorithm}[h]
\caption{Approximation-Aware GeLU Training}
\label{alg:robust_gelu}
\begin{algorithmic}[1]
\Require input $x$, boundary $B$,
coefficients $a,b,c$, sharpness $k$,
noise region $\mathcal R_{\mathrm{noise}}$,
noise magnitude $\eta$
\State $m\gets a|x|^2+b|x|+c+0.5x$
\If{training}
    \State
    $w_{\mathrm{pos}}\gets\sigma(k(x-B))$
    \State
    $w_{\mathrm{neg}}\gets\sigma(k(-x-B))$
    \State
    $y\gets
    w_{\mathrm{pos}}x+
    \max(0,1-w_{\mathrm{pos}}-w_{\mathrm{neg}})m$
    \If{$|x|\in\mathcal R_{\mathrm{noise}}$}
        \State sample
        $\epsilon\sim\mathcal U(-\eta,\eta)$
        and set $y\gets y+\epsilon$
    \EndIf
\Else
    \State
    $y\gets x$ if $x>B$;
    $y\gets m$ if $-B\le x\le B$;
    $y\gets0$ otherwise
\EndIf
\State \Return $y$
\end{algorithmic}
\end{algorithm}

For the representative second-order GeLU configuration,
we use
\begin{equation}
\mathcal R_{\mathrm{noise}}
=
[-2.0,-1.2]\cup[1.2,2.0],
\qquad
\eta=0.09,
\qquad
k=10.
\end{equation}

The adaptive perturbation and soft-boundary mechanisms address
two complementary failure modes: large approximation errors in
specific regions and abrupt error variation around segmented
boundaries.
Together they enlarge the low-degree operating region that can be
used during structured degree sparsification.

% ============================================================
\section{Structured Degree Regularization}
\label{Regularization_Setting}

DegreeSpar learns entry-wise degree reduction and structured
sparse degree states within the same offline optimization process.
This appendix gives the regularization objective and the settings
used in our implementation.

Let $\mathcal G$ denote a collection of computation-aligned groups
and $M_g$ the subset of degree variables associated with group
$g$. We use mixed $\ell_1/\ell_2$ regularization
\begin{equation}
r_{\ell_1/\ell_2}(M)
=
\sum_{g\in\mathcal G}
\|M_g\|_2
\label{eq:apdx_group_reg}
\end{equation}
to shrink aligned groups together.

For token selection we additionally use the differentiable
0--1 surrogate
\begin{equation}
r_{0\text{--}1}(m)
=
\left(m(1-m)\right)^2,
\label{eq:apdx_binary_reg}
\end{equation}
whose derivative is
\begin{equation}
\frac{\partial r_{0\text{--}1}}{\partial m}
=
2m(1-m)(1-2m).
\end{equation}
This term discourages intermediate selection states and pushes
the token-selection variable toward $0$ or $1$.

The resulting optimization objective can be written as
\begin{equation}
\mathcal L
=
\mathcal L_{\mathrm{task}}
+
\lambda_{\mathrm{grp}}
r_{\ell_1/\ell_2}(M)
+
\lambda_{\mathrm{bin}}
\sum_m r_{0\text{--}1}(m),
\label{eq:apdx_total_objective}
\end{equation}
where the same degree representation is used for entry-wise
degree reduction and for the token- and dimension-aligned
structured groups.

In our original configuration,
\begin{equation}
\lambda_{\mathrm{grp}}
=
3\times10^{-4},
\qquad
\lambda_{\mathrm{bin}}
=
1\times10^{-2}.
\end{equation}
The group regularizer progressively lowers degree variables and
encourages computation-aligned groups toward zero, while the
binary surrogate stabilizes discrete token-retention decisions.

Degree sparsification follows a progressive schedule.
Entry-wise regularization first reduces individual nonlinear
evaluation degrees to lower secure computation cost.
As the degrees approach low-order states, computation-aligned
group regularization is introduced to encourage removable
zero structures.

In the 60-epoch training configuration, compression starts at
epoch 30 and continues for up to 30 epochs.
Validation accuracy is monitored throughout the compression
stage.
For the ViT experiments, compression is stopped when the validation accuracy drops by more than $1.0$ percentage points below the uncompressed baseline accuracy. The last checkpoint satisfying this accuracy constraint is retained as the final compressed model.
The resulting structured sparsity and token-retention budgets
are then frozen for deployment.

Importantly, DegreeSpar does not treat nonlinear approximation,
token removal, and model-dimension removal as separately
parameterized compression modules.
Instead, it organizes their computational redundancy within
the same structured degree space.
Entry-wise degree reduction lowers nonlinear evaluation cost,
while computation-aligned degree sparsity, together with its
associated structural parameters, enables removable token-
and dimension-level structures to emerge.

% ============================================================
\section{Variance-Guided CLS Attention}
\label{VCA}

DegreeSpar performs token-aligned degree optimization in a canonical
rank space rather than over absolute token identities.
This requires a significance estimator that provides a reliable
ordering of tokens across different inputs.
Importantly, the significance score is used only to construct this
ordering; it does not directly determine polynomial degrees or
token-removal decisions.
The degree values and structured sparse groups are learned through the
structured degree optimization described in
Section~\ref{sec:structured_sparsification}.

\begin{figure}[h]
\centering
\includegraphics[trim={0cm 0cm 0cm 0cm}, clip, scale=0.24]
{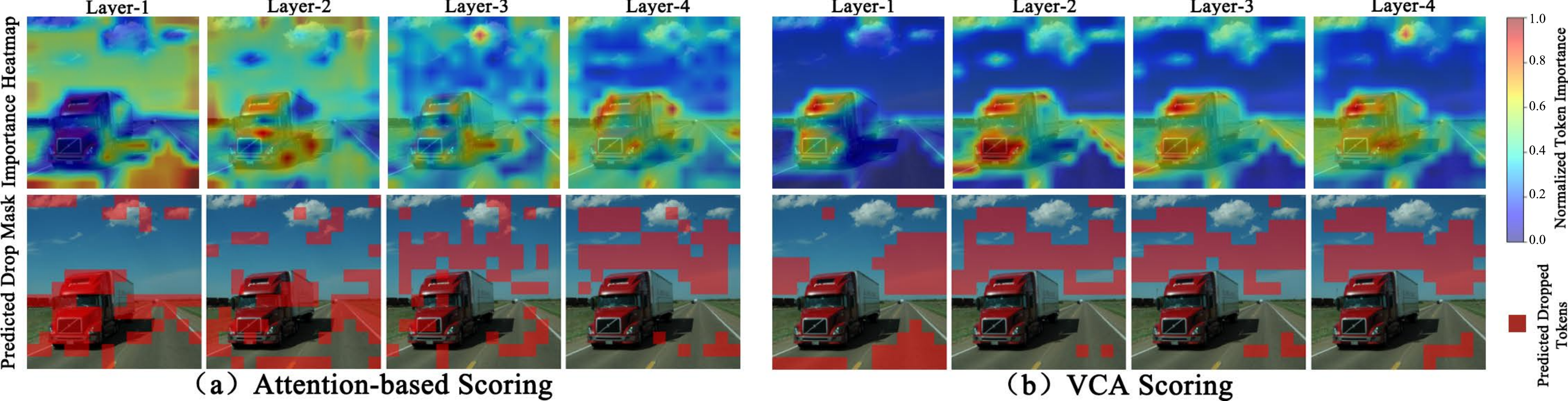}
\vspace*{-0.05in}
\caption{ Comparison between attention-based and VCA token scoring across shallow ViT layers. Red squares indicate tokens predicted to be dropped by each scoring rule at the corresponding layer, without actually removing them. Token-drop predictions are made independently at each layer and are not propagated to subsequent layers. }
\label{fig:vca_cmp}
% \vspace*{-0.10in}
\end{figure}

A natural choice is to rank tokens using their attention scores.
However, attention magnitude does not always provide a reliable
measure of token importance, particularly in shallow Vision
Transformer layers.
As illustrated in Figure~\ref{fig:vca_cmp}, attention-based scoring
can assign high importance to semantically simple or background
patches in early layers, making aggressive early token removal
unreliable.
Introducing an additional prediction network could improve token
importance estimation, but evaluating such a network on encrypted
features would introduce additional secure-inference overhead.

We therefore use \textbf{Variance-Guided CLS Attention (VCA)} as a
secure-inference-friendly ranking estimator.
For token $i$ at layer $l$, VCA combines token-wise feature variance
$V_i^{(l)}$ with the \texttt{[CLS]} attention weight
$A_{\mathrm{cls}\rightarrow i}^{(l)}$:
\begin{equation}
S_i^{(l)}
=
\alpha_l
\phi\!\left(V_i^{(l)}\right)
+
(1-\alpha_l)
\phi\!\left(
A_{\mathrm{cls}\rightarrow i}^{(l)}
\right),
\label{eq:apdx_vca}
\end{equation}
where
\begin{equation}
\alpha_l
=
\frac{1}
{1+\exp(l-L_{\mathrm{trans}})},
\end{equation}
$\phi(\cdot)$ denotes normalization, and $L_{\mathrm{trans}}$
controls the transition between the two signals.

The variance term receives larger weight in shallow layers, where
local feature variation provides additional information for
distinguishing informative patches.
As the network becomes deeper, the weighting gradually shifts toward
\texttt{[CLS]} attention, which increasingly reflects
task-relevant semantic information.
For ViT-S on ImageNet-1K, we use
\begin{equation}
L_{\mathrm{trans}}=4.
\end{equation}

Figure~\ref{fig:vca_cmp} qualitatively illustrates the motivation for this hybrid ranking rule. The red squares indicate tokens predicted to be dropped by each scoring method at the corresponding layer. These predictions are visualized independently at each layer, without actually removing tokens or propagating the predicted removal decisions to subsequent layers. The difference is particularly relevant to DegreeSpar because the ranking defines the canonical token positions over which offline structured degree optimization learns a fixed retention budget. An inaccurate ordering can place informative tokens in low-ranked positions, limiting how aggressively the model can remove tokens without degrading accuracy.

VCA does not introduce an auxiliary prediction module.
The variance term is obtained from statistics already available
around LayerNorm, while the \texttt{[CLS]} attention weights are
produced by the standard attention computation.
During secure inference, these quantities and the resulting
significance scores remain inside the secure-computation domain;
their evaluation uses the same secret-sharing and MPC primitives
as the surrounding Transformer computation.

\begin{figure}[h]
\centering
\includegraphics[trim={0cm 0cm 0cm 0cm}, clip, scale=0.72]
{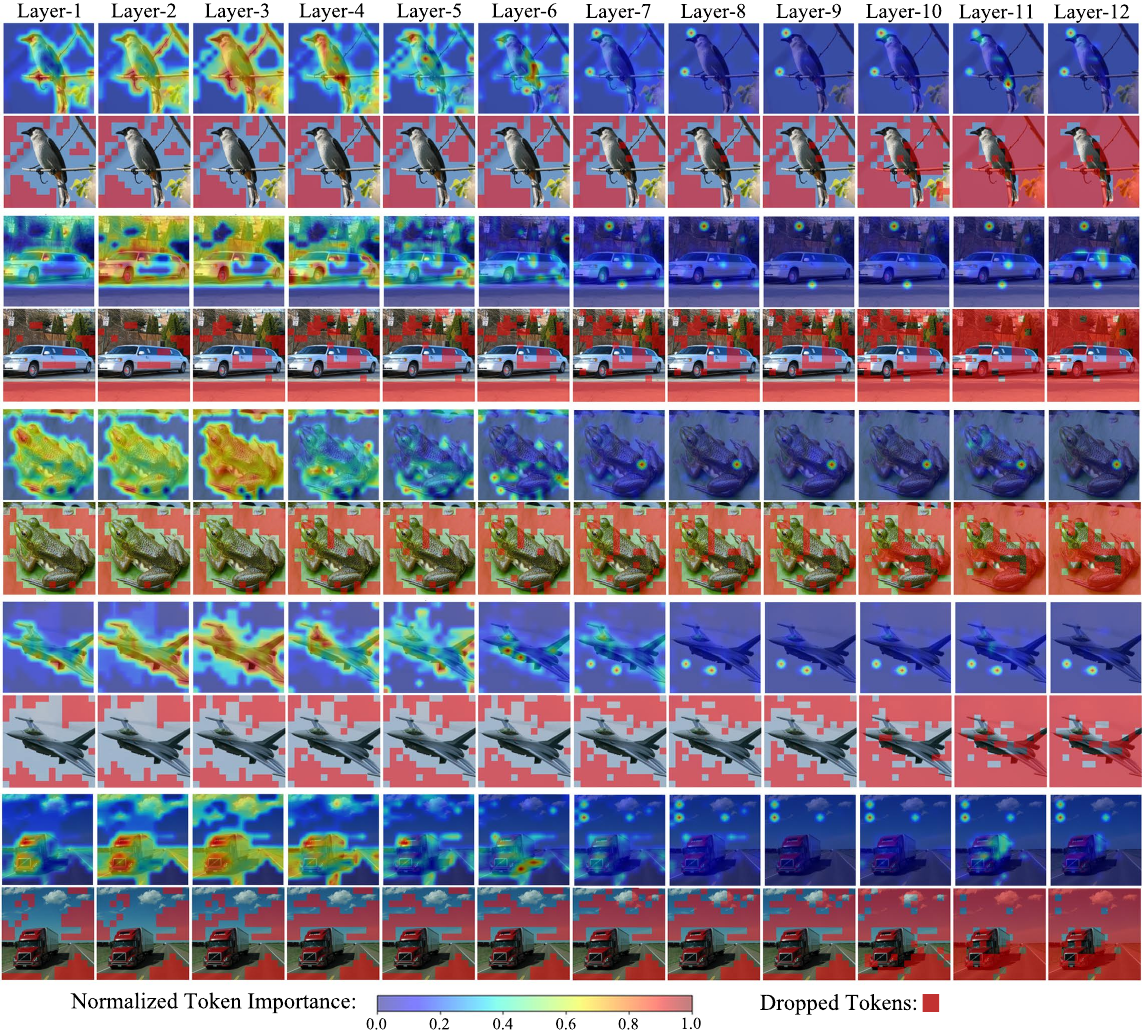}
\vspace*{-0.10in}
\caption{
Visualization of VCA-guided token skipping across Transformer
layers.
The ranking supports token removal from shallow layers while
preserving the fixed per-layer retention budgets learned offline.
}
\label{fig:vca_drop}

\end{figure}

Figure~\ref{fig:vca_drop} provides an additional visualization of
the resulting token-removal patterns.
The results show that the VCA ranking can support token skipping
from shallow layers, including Layer~1, rather than requiring the
method to wait until later layers where attention maps become more
stable.

% ============================================================
\section{Offline Model-Dimension Compaction}
\label{apdx:dimension_compaction}

Dimension-aligned structured sparsity is converted into physical
model compression before secure deployment.

Consider an FFN block
\begin{equation}
h
=
W_2\,\mathrm{GeLU}(W_1x).
\end{equation}
An intermediate dimension $j$ jointly corresponds to an output
dimension of the first projection, one GeLU coordinate, and the
matching input dimension of the second projection.
DegreeSpar groups the GeLU degree variables associated with this
intermediate dimension across the relevant token positions.

The learned dimension-aligned degree sparsity is translated
into physical model compaction through the associated
structural parameters.
Once the resulting zero structure eliminates the contribution
of an intermediate dimension, the corresponding computational
unit is removed jointly.
Under the matrix convention
$W_1\in\mathbb R^{d_{\mathrm{model}}\times D}$ and
$W_2\in\mathbb R^{D\times d_{\mathrm{model}}}$,
this removes
\begin{equation}
W_1[:,j],
\qquad
\mathrm{GeLU}_j,
\qquad
W_2[j,:].
\end{equation}
The remaining dimensions are then compacted into smaller dense
matrices.

The deployed model therefore contains physically smaller dense
FFN layers rather than the original matrices with zero entries or
runtime pruning masks.
No model-dimension selection is performed during private online
inference.
Once the compact model has been constructed offline, the
underlying secure-inference framework evaluates it exactly as a
fixed Transformer with a smaller intermediate dimension.

% ============================================================

% ============================================================
\section{Secure Token Ranking and Removal}
\label{Secure_token_drop}

Unlike model dimensions, the identities of redundant tokens
can vary with the private input.
DegreeSpar therefore determines the token-retention budget
offline while allowing the identities of retained tokens to
remain input dependent during secure inference.

For Transformer layer $l$, let $N_l$ denote the input token
length and $K_l$ denote the retention budget learned during
offline optimization.
The mapping $N_l \rightarrow K_l$ is fixed before deployment
and remains independent of the private input.
During online inference, token significance scores are
evaluated within the secure-computation domain, and an
oblivious top-$K_l$ selection protocol identifies and retains
the corresponding tokens without revealing their identities.

The protocol performs order-preserving top-$K$ extraction
using secure comparisons, oblivious sorting, and arithmetic
routing.
Token features, significance scores, selection thresholds,
and pruning decisions remain secret-shared throughout
execution.
Only the predefined retention budget and the resulting
tensor shape are public.

The following subsections detail the complete secure
token-selection procedure and its underlying oblivious
sorting and routing protocols.

% ============================================================
\subsection{Oblivious Top-$K$ Token Selection}

The oblivious token-selection protocol reduces an input
sequence of length $N$ to a fixed length $K$ according to
the secret-shared token significance scores.
It securely identifies the $K$ highest-ranked tokens,
moves them into a designated partition, and truncates
the remaining sequence without revealing which original
tokens have been retained.

Importantly, the retained tokens preserve their original
relative order rather than being output in significance-score
order.
This property maintains the correspondence between
retained token features and their original sequence
positions.
Algorithm~\ref{alg:mpc_topk_pruning} describes the complete
procedure, which consists of four phases.

\begin{algorithm}[t]
\caption{Oblivious Top-$K$ Token Selection
$\Pi_{\mathsf{Prune}}$}
\label{alg:mpc_topk_pruning}

\textbf{Public Parameters:}
Sequence length $N$, target retention length $K$
($0<K<N$), and feature dimension $D$.

\textbf{Input:}
Secret-shared token significance scores
$\llbracket\mathbf S\rrbracket
\in\mathbb Z_q^N$,
and associated token features
$\llbracket\mathbf V\rrbracket,
\llbracket\mathbf H\rrbracket
\in\mathbb Z_q^{N\times D}$.

\textbf{Output:}
Retained secret-shared features
$\llbracket\mathbf V_p\rrbracket,
\llbracket\mathbf H_p\rrbracket
\in\mathbb Z_q^{K\times D}$,
preserving the original relative order of retained tokens.

\begin{algorithmic}[1]

\Statex \textbf{Phase 1: Oblivious Threshold Extraction}

\State Sort the secret-shared significance scores
in ascending order:
\State
$\displaystyle
\llbracket\mathbf S^*\rrbracket
\gets
\Pi_{\mathsf{Sort}}
(\llbracket\mathbf S\rrbracket)
$

\State Extract the $K$-th largest score as the
secret-shared selection threshold:
\State
$\displaystyle
\llbracket\tau\rrbracket
\gets
\llbracket S^*_{N-K}\rrbracket
$

\Statex \textbf{Phase 2: Order-Preserving Key Generation}

\For{$i=0,\ldots,N-1$}

    \State Securely compare the original significance
    score with the selection threshold:
    \State
    $\displaystyle
    \llbracket c_i\rrbracket
    \gets
    \Pi_{\mathsf{Cmp}}
    (\llbracket S_i\rrbracket,
     \llbracket\tau\rrbracket)
    $
    \Comment{$c_i=1$ for a retained token}

    \State Construct the corresponding routing key:
    \State
    $\displaystyle
    \llbracket k_i\rrbracket
    \gets
    N\llbracket c_i\rrbracket+i
    $

\EndFor

\Statex \textbf{Phase 3: Oblivious Routing}

\State Sort the routing keys in ascending order,
applying the same secret permutation to all
associated token payloads:
\State
$\displaystyle
\begin{aligned}
&(
\llbracket\mathbf K^*\rrbracket,
\llbracket\mathbf V^*\rrbracket,
\llbracket\mathbf H^*\rrbracket
)
\\
&\quad\gets
\Pi_{\mathsf{SortByKey}}
\left(
\llbracket\mathbf K\rrbracket,
\{
\llbracket\mathbf V\rrbracket,
\llbracket\mathbf H\rrbracket
\}
\right)
\end{aligned}
$

\Statex \textbf{Phase 4: Public Truncation}

\State Locally truncate the routed payload arrays
to retain their last $K$ entries:
\State
$\displaystyle
\llbracket\mathbf V_p\rrbracket
\gets
\{
\llbracket V_i^*\rrbracket
\}_{i=N-K}^{N-1}
$
\State
$\displaystyle
\llbracket\mathbf H_p\rrbracket
\gets
\{
\llbracket H_i^*\rrbracket
\}_{i=N-K}^{N-1}
$

\State \Return
$\llbracket\mathbf V_p\rrbracket,
 \llbracket\mathbf H_p\rrbracket$

\end{algorithmic}
\end{algorithm}

\paragraph{Threshold extraction and token selection.}

The protocol first invokes the oblivious sorting functionality
$\Pi_{\mathsf{Sort}}$ to arrange the secret-shared significance
scores in ascending order.
The $K$-th largest score, located at index $N-K$ under
zero-based indexing, is extracted as the selection threshold
$\llbracket\tau\rrbracket$.

Each original score is then securely compared with this
threshold to produce a secret-shared selection bit
$\llbracket c_i\rrbracket$.
A value of $c_i=1$ indicates that token $i$ is selected
for retention, whereas $c_i=0$ indicates that it is removed.

The selection procedure assumes that the cutoff separates
exactly $K$ tokens from the remaining sequence.
Under distinct significance scores, retaining scores greater
than or equal to the $K$-th largest score produces exactly
$K$ selected tokens.
The selection bits remain secret-shared and are never
revealed to either party.

\paragraph{Order-preserving oblivious routing.}

The selection bits are converted into routing keys according to
\begin{equation}
\llbracket k_i\rrbracket
=
N\llbracket c_i\rrbracket+i.
\end{equation}

This construction partitions the routing keys into two
nonoverlapping ranges:
\begin{equation}
k_i\in
\begin{cases}
[0,N-1], & c_i=0,\\
[N,2N-1], & c_i=1.
\end{cases}
\end{equation}

Consequently, sorting the routing keys in ascending order
moves all retained tokens into the final $K$ positions of
the sequence.

Within each partition, the original index $i$ determines
the relative ordering of tokens.
Therefore, the sorting operation preserves the original
relative order of the retained tokens rather than
rearranging them according to their significance scores.

The oblivious sorting functionality
$\Pi_{\mathsf{SortByKey}}$ applies identical secret
permutations to the routing keys and their associated
token features.
Thus, the retained features remain correctly aligned
throughout the routing procedure.

Finally, the protocol locally truncates the routed
secret-shared arrays to their final $K$ positions.
Because the number of retained tokens is fixed offline
and the routing permutation remains secret, this
truncation reveals only the predefined output shape,
not the identities or original positions of the
retained tokens.

% ============================================================
\subsection{Secure Conditional Swap}

Oblivious sorting requires conditionally exchanging
secret-shared values without revealing the comparison
outcome.
A conventional software branch based on a secret
comparison result could expose the underlying ordering
through data-dependent control flow.
To avoid this leakage, we implement conditional
swapping entirely through arithmetic operations
on secret shares.

Given a secret-shared control bit
$\llbracket c\rrbracket\in\{0,1\}$ and operands
$\llbracket x\rrbracket,\llbracket y\rrbracket$,
the oblivious swap protocol
$\Pi_{\mathsf{OSwap}}$ returns the original operands
when $c=0$ and exchanges them when $c=1$.

\begin{algorithm}[h]
\caption{Oblivious Swap Protocol
$\Pi_{\mathsf{OSwap}}$}
\label{alg:oblivious_swap}

\textbf{Input:}
Secret-shared control bit
$\llbracket c\rrbracket\in\{0,1\}$,
and secret-shared operands
$\llbracket x\rrbracket,\llbracket y\rrbracket$.

\textbf{Output:}
Conditionally exchanged operands
$\llbracket x'\rrbracket,
\llbracket y'\rrbracket$.

\begin{algorithmic}[1]

\State Compute the difference using local arithmetic:
\State
$\displaystyle
\llbracket\Delta\rrbracket
\gets
\llbracket x\rrbracket-
\llbracket y\rrbracket
$

\State Securely compute the swap mask
using one interactive multiplication:
\State
$\displaystyle
\llbracket\mu\rrbracket
\gets
\llbracket c\rrbracket
\cdot
\llbracket\Delta\rrbracket
$

\State Apply the mask locally to conditionally
exchange the operands:
\State
$\displaystyle
\llbracket x'\rrbracket
\gets
\llbracket x\rrbracket-
\llbracket\mu\rrbracket
$
\State
$\displaystyle
\llbracket y'\rrbracket
\gets
\llbracket y\rrbracket+
\llbracket\mu\rrbracket
$

\State \Return
$\llbracket x'\rrbracket,
\llbracket y'\rrbracket$

\end{algorithmic}
\end{algorithm}

When $c=0$, the swap mask is zero and both operands
remain unchanged.
When $c=1$, the swap mask equals $x-y$, causing
the two operands to exchange their values.

The difference computation and final operand updates
require only local arithmetic on secret shares,
while computing the swap mask requires one
interactive secure multiplication.
Importantly, the control bit is never reconstructed
or revealed.
The protocol therefore performs conditional
swapping without introducing data-dependent
software branches or revealing the comparison result.

% ============================================================
\subsection{Data-Independent Bitonic Sorting}

DegreeSpar uses oblivious sorting to rank secret-shared
significance scores and route token features according
to secret selection decisions.
We instantiate the sorting functionality using a
Bitonic sorting network.
Unlike conventional sorting algorithms with
data-dependent execution paths, a Bitonic sorting
network follows a predefined sequence of comparisons
and conditional swaps.
The compared index pairs and sorting directions are
determined entirely by the public sequence length
and the sorting-network structure.

Algorithm~\ref{alg:bitonic_sort} describes the
oblivious sorting procedure.
The outer loop progressively constructs bitonic
sequences, while the inner loop merges them through
compare-and-swap operations at decreasing strides.

\begin{algorithm}[h]
\caption{Oblivious Bitonic Sort Protocol
$\Pi_{\mathsf{Sort}}$}
\label{alg:bitonic_sort}

\textbf{Input:}
Secret-shared array
$\llbracket\mathbf X\rrbracket
\in\mathbb Z_q^N$.

\textbf{Output:}
Secret-shared array
$\llbracket\mathbf X^*\rrbracket$
sorted in ascending order.

\begin{algorithmic}[1]

\For{$k\in\{2^1,2^2,\ldots,N\}$}
    \Comment{Build bitonic sequences of size $k$}

    \For{$d\in\{k/2,k/4,\ldots,1\}$}
        \Comment{Merge sequences with stride $d$}

        \State Let $\mathcal E_{k,d}$ be the fixed
        set of valid index pairs $(u,v)$
        separated by stride $d$.

        \For{each $(u,v)\in\mathcal E_{k,d}$}

            \State Determine the ascending or
            descending comparison direction
            from the public sorting-network structure.

            \State Securely compare the two elements
            according to the required direction:
            \State
            $\displaystyle
            \llbracket c\rrbracket
            \gets
            \Pi_{\mathsf{Cmp}}
            (\llbracket x_u\rrbracket,
             \llbracket x_v\rrbracket)
            $

            \State Conditionally exchange the elements
            according to the secret comparison bit:
            \State
            $\displaystyle
            \begin{aligned}
            &(
            \llbracket x_u\rrbracket,
            \llbracket x_v\rrbracket
            )
            \\
            &\quad\gets
            \Pi_{\mathsf{OSwap}}
            (
            \llbracket c\rrbracket,
            \llbracket x_u\rrbracket,
            \llbracket x_v\rrbracket
            )
            \end{aligned}
            $

        \EndFor

    \EndFor

\EndFor

\State \Return $\llbracket\mathbf X^*\rrbracket$

\end{algorithmic}
\end{algorithm}

The sorting network uses a fixed sequence of comparison
pairs, and the ascending or descending direction of
each comparison is determined by public indices.
Neither the comparison pairs nor the execution order
depends on the secret-shared input values.
Each comparison produces a secret-shared control bit
that determines whether the corresponding elements
should be exchanged.
The control bit is passed directly to
$\Pi_{\mathsf{OSwap}}$ without being revealed.

Consequently, the sorting procedure evaluates a
data-independent sequence of secure comparisons
and arithmetic swaps.
The underlying ranking and comparison outcomes remain
protected throughout execution.

\paragraph{Oblivious sorting with associated payloads.}

Sorting significance scores or routing keys alone is
insufficient for token removal.
The corresponding token features must undergo exactly
the same permutation so that each feature vector remains
associated with its original significance score and
selection decision.

We therefore use the payload-aware sorting functionality
$\Pi_{\mathsf{SortByKey}}$, described in
Algorithm~\ref{alg:bitonic_sort_payload}.
For every comparison pair, the protocol first securely
compares the routing keys and obtains a secret-shared
control bit.
The same control bit is then applied to both the keys
and all associated payloads.
This ensures that the payloads follow the key permutation
without revealing the underlying token identities.

\begin{algorithm}[h]
\caption{Oblivious Bitonic Sort with Payload
$\Pi_{\mathsf{SortByKey}}$}
\label{alg:bitonic_sort_payload}

\textbf{Input:}
Secret-shared routing keys
$\llbracket\mathbf K\rrbracket
\in\mathbb Z_q^N$,
and an optional set of associated secret-shared
payloads
$\mathcal P=
\{
\llbracket\mathbf P^{(1)}\rrbracket,
\llbracket\mathbf P^{(2)}\rrbracket,
\ldots
\}$.

\textbf{Output:}
Sorted secret-shared routing keys
$\llbracket\mathbf K^*\rrbracket$
and correspondingly permuted payloads
$\mathcal P^*$.

\begin{algorithmic}[1]

\For{$k\in\{2^1,2^2,\ldots,N\}$}
    \Comment{Build bitonic sequences of size $k$}

    \For{$d\in\{k/2,k/4,\ldots,1\}$}
        \Comment{Merge sequences with stride $d$}

        \State Let $\mathcal E_{k,d}$ be the fixed
        set of valid index pairs $(u,v)$
        separated by stride $d$.

        \For{each $(u,v)\in\mathcal E_{k,d}$}

            \State Determine the ascending or
            descending comparison direction
            from the public sorting-network structure.

            \State Securely compare the routing keys
            according to the required direction:
            \State
            $\displaystyle
            \llbracket c\rrbracket
            \gets
            \Pi_{\mathsf{Cmp}}
            (
            \llbracket k_u\rrbracket,
            \llbracket k_v\rrbracket
            )
            $

            \State Conditionally exchange the keys:
            \State
            $\displaystyle
            \begin{aligned}
            &(
            \llbracket k_u\rrbracket,
            \llbracket k_v\rrbracket
            )
            \\
            &\quad\gets
            \Pi_{\mathsf{OSwap}}
            (
            \llbracket c\rrbracket,
            \llbracket k_u\rrbracket,
            \llbracket k_v\rrbracket
            )
            \end{aligned}
            $

            \If{$\mathcal P\neq\emptyset$}
                \Comment{Apply identical swaps to all payloads}

                \For{each
                $\llbracket\mathbf P\rrbracket
                \in\mathcal P$}

                    \State
                    $\displaystyle
                    \begin{aligned}
                    &(
                    \llbracket p_u\rrbracket,
                    \llbracket p_v\rrbracket
                    )
                    \\
                    &\quad\gets
                    \Pi_{\mathsf{OSwap}}
                    (
                    \llbracket c\rrbracket,
                    \llbracket p_u\rrbracket,
                    \llbracket p_v\rrbracket
                    )
                    \end{aligned}
                    $

                \EndFor

            \EndIf

        \EndFor

    \EndFor

\EndFor

\State \Return
$\llbracket\mathbf K^*\rrbracket,\mathcal P^*$

\end{algorithmic}
\end{algorithm}

The payload-aware sorting protocol differs from
the ordinary sorting functionality only in the
synchronized exchange of associated payloads.
Each secret-shared comparison bit is reused across
the corresponding routing keys and all payload
arrays.
For vector-valued payloads, the conditional swap
is applied to every feature coordinate under the
same control bit.
Thus, the resulting permutation preserves the
alignment between routing keys, token features,
and any additional associated secret-shared values.
When combined with the order-preserving routing-key
construction in Algorithm~\ref{alg:mpc_topk_pruning},
this protocol places the selected tokens in the
retained partition while preserving their original
relative order.

% ============================================================
\paragraph{Security and observable information.}

The complete token-selection procedure consists of
secure threshold extraction, secret routing-key
construction, oblivious payload routing, and public
truncation.

All significance scores, intermediate comparisons,
selection bits, routing keys, and token features
remain secret-shared throughout the procedure.
The sorting-network comparison schedule is fixed
by public parameters, and conditional swaps are
performed without data-dependent branches.
Although the selected token identities depend on
the private input, the number of retained tokens
is determined by the fixed offline retention budget.
The final truncation therefore exposes only the
predefined output shape.

Under the two-party semi-honest threat model and
the security guarantees of the underlying MPC
primitives, neither party learns the significance
scores, ranking outcomes, or identities of retained
and removed tokens from the token-selection protocol.
The complete threat model and the corresponding
security analysis are provided in
Appendix~\ref{security_guarantee}.

% ============================================================
\section{Cryptographic Preliminaries and Security Analysis}
\label{security_guarantee}

This section summarizes the cryptographic primitives underlying
DegreeSpar and analyzes the information revealed by its compression
mechanisms.
DegreeSpar does not introduce a new cryptographic protocol.
Instead, it operates on top of the standard two-party secure
Transformer inference setting adopted by existing systems such as
BOLT and BumbleBee~\citep{pang2023bolt,lu2023bumblebee}.
Linear computation inherits the protection of the underlying
HE-based inference framework, while nonlinear and data-dependent
operations are implemented using standard secure multi-party
computation primitives.
The complete oblivious token-selection protocols used by DegreeSpar
are given separately in Appendix~\ref{Secure_token_drop}.

\subsection{Cryptographic Preliminaries}

\paragraph{Homomorphic encryption.}
Homomorphic encryption (HE) enables arithmetic operations to be
evaluated directly over encrypted data without first revealing the
underlying plaintext.
Let
\[
    c = \mathsf{Enc}_{pk}(x)
\]
denote the encryption of a plaintext value $x$ under public key
$pk$.
An HE scheme provides ciphertext-domain operations whose decrypted
results correspond to arithmetic over the plaintext values.
For example, for supported operations $\oplus$ and $\otimes$,
\[
\mathsf{Dec}_{sk}
\big(
    \mathsf{Enc}_{pk}(x)
    \oplus
    \mathsf{Enc}_{pk}(y)
\big)
=
x+y,
\]
and, when ciphertext multiplication is supported,
\[
\mathsf{Dec}_{sk}
\big(
    \mathsf{Enc}_{pk}(x)
    \otimes
    \mathsf{Enc}_{pk}(y)
\big)
=
xy.
\]

Secure Transformer frameworks exploit these properties to evaluate
linear operations such as matrix multiplication over protected
activations.
DegreeSpar does not modify the HE protocol used by the underlying
framework.
Its offline dimension compaction simply reduces the dimensions of
the linear operators that must subsequently be evaluated securely.

\paragraph{Additive secret sharing.}
Interactive nonlinear and data-dependent operations are represented
using additive secret sharing.
A value
$x\in\mathbb Z_{2^{\mathcal L}}$
is represented by two shares held by parties $P_0$ and $P_1$:
\[
\llbracket x\rrbracket
=
\left(
\langle x\rangle_0,
\langle x\rangle_1
\right),
\]
such that
\[
x
=
\langle x\rangle_0
+
\langle x\rangle_1
\pmod{2^{\mathcal L}}.
\]
Neither share alone reveals the underlying value.

Addition of two secret-shared values is local:
\[
\llbracket x+y\rrbracket
=
\left(
\langle x\rangle_0+\langle y\rangle_0,
\langle x\rangle_1+\langle y\rangle_1
\right).
\]
More complex operations, including multiplication and comparison,
require interaction between the two parties and are instantiated
using the secure primitives of the underlying MPC framework.

\paragraph{Oblivious transfer.}
Oblivious transfer (OT) is a standard building block for secure
two-party computation.
In a 1-out-of-2 OT, the sender holds two messages
$m_0$ and $m_1$, while the receiver holds a private choice bit
$c\in\{0,1\}$.
At the end of the protocol, the receiver obtains $m_c$ without
learning the other message, while the sender learns nothing about
$c$.
OT and its extensions are commonly used to implement secure
nonlinear operations, multiplication, comparison, and conditional
selection in two-party secure inference systems
~\citep{brassard1986all,asharov2013more,demmler2015aby,
sureshaby2,mohassel2018aby3}.

\paragraph{Secure comparison and multiplexing.}
DegreeSpar requires comparisons when ranking token-significance
scores.
Given secret-shared inputs
$\llbracket x\rrbracket$ and $\llbracket y\rrbracket$,
a secure comparison protocol produces a secret-shared bit
\[
\llbracket c\rrbracket
=
\Pi_{\mathsf{Cmp}}
(
\llbracket x\rrbracket,
\llbracket y\rrbracket
),
\]
where $c$ indicates the comparison result but is not revealed to
either party.

A secret comparison result can subsequently control a conditional
selection without introducing a plaintext branch.
For example, a secure multiplexer can compute
\[
\llbracket z\rrbracket
=
\llbracket c\rrbracket
\llbracket x\rrbracket
+
(1-\llbracket c\rrbracket)
\llbracket y\rrbracket.
\]
Because the control bit remains secret shared, neither the selected
branch nor the comparison outcome is exposed.

\paragraph{Oblivious sorting and routing.}
A conventional sorting procedure would expose information through
data-dependent comparisons, branches, or memory accesses.
DegreeSpar therefore uses a sorting network whose comparison pattern
is determined only by the public sequence length.
In particular, our implementation uses Bitonic sorting together
with secure comparison and conditional swap.
The compared index pairs are fixed before execution.
The comparison outcomes remain secret shared, and the same secret
swap bits are applied to the associated token payloads.
Consequently, an observer sees the same sorting-network structure
independent of the private significance scores.
Appendix~\ref{Secure_token_drop} gives the complete secure
comparison, conditional-swap, Bitonic-sort, payload-routing, and
Top-$K$ procedures used by DegreeSpar.

\subsection{Threat Model}

DegreeSpar follows the standard two-party semi-honest adversary
model used by the underlying secure Transformer inference
frameworks~\citep{lu2023bumblebee,pang2023bolt,zhang2024secure,
juvekar2018gazelle,rathee2020cryptflow2}.
The two parties follow the prescribed protocol correctly but may
attempt to infer additional information from the messages and
intermediate values available to them during execution.

We consider a client that provides the private Transformer input
and a server that holds the deployed model parameters.
Following the common threat model of existing secure-inference
systems, the model architecture and the public execution
configuration are assumed to be known.
The cryptographic protocol protects private client inputs and
intermediate activations while preventing either party from
learning secret intermediate values beyond the prescribed output.

DegreeSpar does not change the cryptographic assumptions of the
underlying secure-inference framework.
Its additional security consideration arises from compression:
an input-dependent compression mechanism could reveal information
if the number or positions of retained computations became
externally observable.
We therefore distinguish carefully between compression decisions
made offline and input-dependent decisions made during secure
online inference.

\subsection{Offline Compression versus Online Private Selection}

DegreeSpar learns its compression policy entirely before secure
deployment.
Polynomial-degree structures, model-dimension sparsity, and the
number of token positions retained in each Transformer layer are
determined during offline optimization.

\paragraph{Polynomial-degree configuration.}
Degree reduction changes the nonlinear functions executed by the
deployed model but does not introduce an input-dependent online
control decision.
After training, the learned degree configuration is fixed during
secure inference.
The corresponding nonlinear functions are then evaluated using the
same secure primitives as in the underlying inference framework.

\paragraph{Dimension-aligned sparsity.}
Dimension-aligned sparse groups are also resolved completely
offline.
As described in Appendix~\ref{apdx:dimension_compaction}, the
corresponding FFN dimensions are physically removed and the
surrounding linear weights are compacted before deployment.
Online secure inference therefore evaluates a fixed compacted
network.
No private input determines whether a model dimension is retained
or removed at runtime.

\paragraph{Token-aligned sparsity.}
Token-level sparsity differs because the identities of unimportant
tokens depend on the private input.
DegreeSpar therefore separates the \emph{compression budget} from
the \emph{token identities}.

For layer $l$, offline optimization determines a fixed retention
budget
\[
N_l \rightarrow K_l.
\]
The value $K_l$ is fixed before deployment and is shared by all
inputs.
During online inference, however, different inputs may assign
different private token identities to the top-$K_l$ rank
positions.
These identities are determined entirely within secure computation
through the protocol in Appendix~\ref{Secure_token_drop}.

This distinction is central to the security of DegreeSpar:
\[
\underbrace{
K_l
}_{\text{public, fixed, input independent}}
\qquad\text{versus}\qquad
\underbrace{
\text{scores, ranking, and selected identities}
}_{\text{input dependent but hidden}} .
\]

\subsection{Security of Online Token Selection}

At layer $l$, let
\[
\llbracket
\mathbf S^{(l)}
\rrbracket
=
(
\llbracket S_1^{(l)}\rrbracket,\ldots,
\llbracket S_{N_l}^{(l)}\rrbracket
)
\]
denote the secret-shared significance scores.
The online token-selection protocol performs four conceptual steps.

First, the scores are sorted using an oblivious sorting network.
Because the sequence of compared positions depends only on the
public length $N_l$, the sorting control flow is independent of the
private scores.

Second, the Top-$K_l$ threshold
$\llbracket\tau_l\rrbracket$
is extracted while remaining secret shared.
Each original score is securely compared with this threshold,
producing secret-shared keep/drop bits
$\llbracket c_i\rrbracket$.

Third, these secret bits are converted into routing keys and used
to obliviously move retained tokens into a common partition.
The same hidden permutation is applied to the corresponding token
features and other required payloads.

Finally, after the retained tokens have been moved into the
designated partition, the tensor is truncated to the public length
$K_l$.
At this stage, revealing the output shape exposes only the
predefined retention budget.
It does not reveal which original token identities occupy the
retained positions.

Thus, the token-selection protocol keeps private:

\begin{itemize}[leftmargin=0.18in]
    \item token features and intermediate representations;
    \item token-significance scores;
    \item the Top-$K_l$ threshold;
    \item pairwise comparison outcomes;
    \item keep/drop decision bits;
    \item the permutation induced by secure sorting and routing;
    \item the original indices of retained and removed tokens.
\end{itemize}

Only the predefined retention budget and resulting tensor shape
are externally visible.

\subsection{Observable and Protected Information}

\begin{table}[h]
\centering
\caption{
Observable and protected information during DegreeSpar secure
execution.
}
\label{tab:security_observable}
\small
\begin{tabular}{lc}
\toprule
\textbf{Information} & \textbf{Status} \\
\midrule
Per-layer retention/removal budget $K_l$
& Public and fixed \\
Retained tensor shape
& Public and fixed \\
Model architecture under the assumed threat model
& Public and fixed\\
\midrule
Client input
& Hidden \\
Token values/features
& Hidden \\
Token-significance scores
& Hidden \\
Top-$K$ threshold
& Hidden \\
Pairwise comparison outcomes
& Hidden \\
Token movements during sorting
& Hidden \\
Keep/drop decision bits
& Hidden \\
Final selection indices
& Hidden \\
Retained/removed token identities
& Hidden \\
\bottomrule
\end{tabular}
\end{table}

Table~\ref{tab:security_observable} summarizes the resulting
information boundary.

Because $K_l$ is fixed independently of the private input, the
externally observable tensor length does not encode an
input-dependent compression pattern.
Different inputs may retain different tokens, but this variation is
confined to protected values inside the MPC execution.
Consequently, the observable execution shape is identical for all
inputs using the same deployed DegreeSpar configuration.

\subsection{Security of VCA Scoring}

The proposed Variance-Guided CLS Attention (VCA) estimator does not
change this information boundary.
VCA constructs each token-significance score from two quantities
already produced within Transformer inference: token-wise feature
variance derived from LayerNorm-related statistics and
\texttt{[CLS]} attention values.

During secure execution, these intermediate values remain
protected.
Their normalization, weighted combination, comparison, and ranking
are performed using the same secure arithmetic, comparison, and
multiplexer primitives used by the surrounding inference protocol.
The resulting VCA scores are never revealed in plaintext.
They are consumed directly by the oblivious token-selection
procedure.

Therefore, VCA changes how DegreeSpar computes the private ranking
signal, but does not expose the ranking signal or the resulting
token identities.

\subsection{Security Argument for DegreeSpar}

The additional operations introduced by DegreeSpar can be separated
into two classes.

\paragraph{Offline operations.}
Degree optimization, structured regularization, determination of
retention budgets, and model-dimension compaction are completed
before private online inference.
They therefore introduce no input-dependent observation during
deployment.

\paragraph{Online operations.}
Input-dependent token ranking and selection are evaluated entirely
using the secure primitives of the underlying two-party framework.
Scores, comparison results, routing decisions, and selected token
identities remain secret shared.
Only the fixed retention budget $K_l$ and corresponding tensor
shape are public.

DegreeSpar therefore does not require revealing an additional
input-dependent compression decision.
Its linear and nonlinear computation inherits the security of the
underlying secure-inference protocols, while its new
input-dependent token-selection operations are implemented using
secure comparison, oblivious sorting, and oblivious routing as
specified in Appendix~\ref{Secure_token_drop}.

Under the assumed semi-honest threat model, DegreeSpar consequently
reveals no additional input-dependent compression information
beyond the public, fixed execution configuration of the underlying
secure-inference system.
Security against malicious parties and attacks outside the adopted
semi-honest model are outside the scope of this work.

\section{Training and Compression Setup}
\label{Training_Setting}

\subsection{Training Configuration}

DegreeSpar starts from pretrained Transformer models and performs offline fine-tuning before secure deployment.
All training experiments are conducted on NVIDIA A100 GPUs.

For the original ViT configurations, we fine-tune the pretrained models for up to 60 epochs using AdamW with a batch size of 48, an initial learning rate of $10^{-5}$, weight decay of $0.01$, and cosine learning-rate decay.

The first 30 epochs adapt the model to low-degree nonlinear approximations.
Structured degree sparsification is introduced at epoch 30 and continues for up to 30 additional epochs.
During this stage, staged degree reduction and computation-aligned group regularization jointly optimize the nonlinear degree structure.

The group-regularization and token-binarization coefficients in the original configuration are
\begin{equation}
\lambda_{\mathrm{grp}}=3\times10^{-4},
\qquad
\lambda_{\mathrm{bin}}=10^{-2}.
\end{equation}

The complete regularization objective and optimization mechanisms are described in Appendix~\ref{Regularization_Setting}.

\subsection{Compression Budget Selection}

DegreeSpar determines its compression configuration through progressive degree sparsification under a predefined validation-performance constraint.

For the ViT experiments, compression is stopped when the validation accuracy drops by more than $1.0$ percentage points below the corresponding uncompressed baseline accuracy. The last checkpoint satisfying this constraint is selected as the final compressed model. NLP task configurations use $1.5$ percentage points as predefined accuracy tolerances.

The resulting degree configuration, per-layer token-retention budgets, and compacted model dimensions are then fixed for secure deployment.

\subsection{Offline Training Cost}

Producing a compressed DegreeSpar model requires a one-time offline optimization process.
In our original experiments, training takes approximately 1--6 hours on NVIDIA A100 GPUs, depending on the model and dataset.

Degree optimization, approximation-aware training, and model-dimension compaction are completed before deployment.
The adaptive noise injection and soft-boundary mechanisms are disabled after training and introduce no additional online secure-inference computation.

During secure inference, DegreeSpar directly evaluates the learned degree configuration and compacted model under the fixed per-layer token-retention budgets.

\section{Baseline Configurations and Comparison Protocols}
\label{apdx:baseline_comparison}

We reproduce existing compression methods using their published configurations whenever applicable and evaluate them under the same hardware, security parameters, network conditions, and secure-inference backend as DegreeSpar. All reported end-to-end latencies include the computation required by the corresponding compression and inference procedures.

\subsection{Comparison with Existing Compression Methods}

\paragraph{Model pruning and token pruning.}
We compare DegreeSpar with WDPruning~\citep{yu2022width} and EViT~\citep{liang2022not}, which provide model and dataset configurations comparable to our ViT evaluation.
We reproduce their published hyperparameter settings and compression configurations whenever applicable, preserving their original compression strengths.
The resulting models are evaluated under the same secure-inference environment as DegreeSpar.

\paragraph{Nonlinear approximation.}
We additionally compare DegreeSpar with PowerSoftmax~\citep{zimerman2024power}.
Since the original work does not provide a directly matching secure-inference configuration for our ViT evaluation, we implement its published approximation protocol on the same model and dataset used by DegreeSpar.
The resulting model is evaluated using the same secure-inference backend and network settings.
This comparison therefore evaluates the published approximation approach under a common execution environment rather than directly comparing latency measurements obtained from different experimental platforms.

\subsection{Comparison with CipherPrune}

CipherPrune~\citep{zhang2025cipherprune} is the closest prior hybrid approach, combining token pruning with polynomial approximation.
We reproduce CipherPrune using BERT-Base and SST-2, following its published model configuration and compression settings.
The reproduced model achieves 92.66\% accuracy, consistent with the result reported in the original work.
We evaluate CipherPrune and DegreeSpar under identical hardware, security, network, and end-to-end latency-accounting settings, including the costs of their respective token-selection procedures.

We consider two operating points for CipherPrune.

\textbf{Original operating point.}
We retain CipherPrune's published compression configuration to preserve its reported accuracy--efficiency trade-off.
This comparison measures the performance of DegreeSpar relative to CipherPrune under its original operating point.

\textbf{Latency-matched operating point.}
We additionally adjust CipherPrune's compression strength to obtain an inference latency comparable to DegreeSpar.
We denote this configuration as CipherPrune*.
This additional comparison examines the resulting model accuracy when the two approaches operate under approximately the same inference latency.

The original operating point is used for the primary comparison, while the latency-matched configuration provides a complementary comparison of the accuracy--latency trade-off.

\subsection{Modular Compression Baseline}

To evaluate the benefit of unified degree-space optimization, we construct a modular baseline that sequentially combines existing nonlinear approximation~\citep{zimerman2024power}, token pruning~\citep{liang2022not}, and model pruning~\citep{yu2022width}.

We retain the published hyperparameter settings of the constituent methods, except for the parameters controlling compression strength.
Specifically, we adjust their nonlinear approximation settings, token-retention budgets, and model-dimension pruning ratios to match the corresponding compression levels achieved by DegreeSpar.
The three compression mechanisms are then applied sequentially to the same model without jointly optimizing their combined effects through a shared degree-space objective.

This component-wise matched configuration ensures that the comparison uses comparable compression levels for all three components, rather than independently selected compression ratios.
Although the individual methods can preserve high model accuracy when applied separately, their combined approximation and pruning perturbations may accumulate.
Moreover, compression decisions optimized separately can act on overlapping computational redundancy without accounting for their combined effects during optimization.

As reported in Table~\ref{tab:core_validation}(b), the modular combination reduces ViT-S accuracy from 80.20\% to 76.41\%, whereas DegreeSpar retains 80.12\%.
This comparison examines whether independently combining existing compression strategies can reproduce the accuracy achieved by unified degree-space optimization under matched component-wise compression levels.

\section{Additional Experimental Analysis}
\label{apdx:robustness}

\subsection{Training Stability}

We evaluate the stability of DegreeSpar across five independent training runs on ViT-Small/ImageNet-1K, ViT-Base/ImageNet-1K, and BERT-Base/IMDB.
Table~\ref{tab:seed_stability} reports the mean and standard deviation of model accuracy across the five runs, whereas the main end-to-end evaluation reports the accuracy and latency of individual model configurations.
Both the uncompressed baselines and DegreeSpar exhibit standard deviations of 0.1--0.2 percentage points across the evaluated settings, demonstrating stable model accuracy across independent training runs.

\begin{table}[h]
\centering
\caption{Accuracy stability across five independent training runs. Results are reported as mean $\pm$ standard deviation (\%).}
\vspace*{0.05in}
\label{tab:seed_stability}
\small
\begin{tabular}{lccc}
\toprule
\textbf{Method}
& \textbf{ViT-S / ImageNet}
& \textbf{ViT-B / ImageNet}
& \textbf{BERT / IMDB} \\
\midrule
Baseline
& $80.09\pm0.2$
& $81.45\pm0.1$
& $89.78\pm0.2$ \\

DegreeSpar
& $80.31\pm0.2$
& $81.05\pm0.2$
& $88.57\pm0.2$ \\
\bottomrule
\end{tabular}
\end{table}

\subsection{Sensitivity to Regularization Strength}

We investigate the sensitivity of DegreeSpar to the structured group-regularization coefficient $\lambda_{\mathrm{grp}}$.

Table~\ref{tab:lambda_sensitivity} reports the accuracy and secure-inference latency of ViT-Small on ImageNet-1K under WAN2 for different regularization strengths.
Increasing the regularization coefficient encourages more aggressive degree sparsification and reduces inference latency.
However, excessively strong regularization introduces larger accuracy degradation.
At the default coefficient of $3\times10^{-4}$, DegreeSpar achieves 80.12\% accuracy with an inference latency of 132.02\,s.
Reducing the coefficient to $1.5\times10^{-4}$ or $3\times10^{-5}$ yields accuracies of 80.10\% and 80.17\%, respectively, while increasing latency.
These results demonstrate that the regularization coefficient controls the accuracy--latency trade-off.
The default setting provides substantial inference acceleration while preserving baseline-level accuracy.

\begin{table}[h]
\centering
\caption{Sensitivity to the structured group-regularization coefficient on ViT-S/ImageNet-1K under WAN2.}
\vspace*{0.05in}
\label{tab:lambda_sensitivity}
\small
\begin{tabular}{lcc}
\toprule
$\boldsymbol{\lambda_{\mathrm{grp}}}$
& \textbf{Latency (s)}
& \textbf{Accuracy (\%)} \\
\midrule
$3\times10^{-3}$
& 63.71 & 76.95 \\

$6\times10^{-4}$
& 111.29 & 79.20 \\

$3\times10^{-4}$ (default)
& 132.02 & 80.12 \\

$1.5\times10^{-4}$
& 207.65 & 80.10 \\

$3\times10^{-5}$
& 310.04 & 80.17 \\
\bottomrule
\end{tabular}
\end{table}

\subsection{Sensitivity to the VCA Transition Layer}

The Variance-Guided CLS Attention (VCA) estimator combines token-wise feature variance with \texttt{[CLS]} attention using a layer-dependent weighting coefficient.
We examine the sensitivity of DegreeSpar to the transition layer $L_{\mathrm{trans}}$, which controls the shift from variance-based scoring toward attention-based scoring.
Table~\ref{tab:vca_sensitivity} reports the resulting ViT-Small accuracy on ImageNet-1K.
The default transition layer $L_{\mathrm{trans}}=4$ achieves 80.12\% accuracy.
Transition layers 5 and 6 yield comparable accuracies of 80.09\% and 79.94\%, respectively.
Earlier transitions at Layers 2 and 3 result in lower accuracy.
These results indicate that DegreeSpar is relatively insensitive to the transition layer around its default setting, while shifting toward attention-based scoring too early can degrade model performance.

\begin{table}[h]
\centering
\caption{Sensitivity to the VCA transition layer on ViT-S/ImageNet-1K.}
\vspace*{0.05in}
\label{tab:vca_sensitivity}
\small
\begin{tabular}{cc}
\toprule
\textbf{Transition Layer}
& \textbf{Accuracy (\%)} \\
\midrule
2 & 78.85 \\
3 & 79.80 \\
4 (default) & 80.12 \\
5 & 80.09 \\
6 & 79.94 \\
\bottomrule
\end{tabular}
\end{table}

\subsection{Secure Token-Selection Overhead}

DegreeSpar performs input-dependent token selection through oblivious ranking and routing under a fixed per-layer retention budget.
Although these operations introduce additional secure computation, their overhead is small relative to the overall secure-inference latency in the evaluated configuration.
For ViT-Small on ImageNet-1K with 197 input tokens under WAN2, token-selection procedure requires 1.73\,s.
This corresponds to approximately 0.32\% of the uncompressed inference latency of 538.52\,s.
Importantly, the secure token-selection cost is already included in the reported end-to-end inference latency.
The performance gains of DegreeSpar therefore account for the additional online cost of privately selecting and removing redundant tokens.
Model-dimension compaction is performed offline before deployment and introduces no additional online dimension-selection procedure.

\end{document}